\documentclass{JHEP3}
\usepackage{cite}
\usepackage{epsfig}
\usepackage{booktabs}

\def\as{\alpha_s}

\def\lb#1{\if 1#1 \ln\beta \else \ln^#1\beta \fi}
\def\lt#1{\if 1#1 \ln 2 \else \ln^#1 2 \fi}

\newcommand{\be}{\begin{equation}}
\newcommand{\ee}{\end{equation}}
\newcommand{\ba}{\begin{eqnarray}}
\newcommand{\ea}{\end{eqnarray}}

\newcommand{\epem}{e^+e^-}
\newcommand{\qb}{\bar{q}}

\newcommand{\qp}{q^\prime}
\newcommand{\qpb}{\bar{q}^\prime}
\newcommand{\qpp}{q^{\prime\prime}}
\newcommand{\qppb}{\bar{q}^{\prime\prime}}

\newcommand{\lsim}{\mbox{\raisebox{-0.3ex}{%
\footnotesize $\:\stackrel{<}{\sim}\:$}} }

\newcommand{\dasi}{\delta\alpha_s^i}
\newcommand{\dasj}{\delta\alpha_s^j}

\title{
NLO  QCD corrections to five-jet production at LEP and the 
extraction of $\alpha_s(M_Z)$
}

\author{Rikkert Frederix\\
Institut f\"ur Theoretische Physik,
Universit\"at Z\"urich,
Winterthurerstrasse 190,\\ 8057 Z\"urich, Switzerland}
\author{Stefano Frixione%
  \thanks{On leave of absence from INFN, Sez. di Genova, Italy.}\\
PH Department, TH Unit, CERN, CH-1211 Geneva 23, Switzerland \\
ITPP, EPFL, CH-1015 Lausanne, Switzerland}
\author{Kirill Melnikov\\
Department of Physics and Astronomy,
Johns Hopkins University,
Baltimore, MD, USA}
\author{ Giulia  Zanderighi\\
Rudolf Peierls Centre for Theoretical Physics,
1 Keble Road, University of Oxford, UK}

\preprint{
 CERN-TH/2010-185\\
 OUTP-1019P\\
 ZU-TH 11/10}

\abstract{ The highest exclusive jet multiplicity studied at LEP
  experiments is five.  In this paper we compute the next-to-leading
  order QCD corrections to $e^+e^-$ annihilation to five jets,
  essentially closing the (pure) perturbative QCD studies of exclusive
  jetty final states at LEP.  We compare fixed-order perturbative
  results with ALEPH data. We estimate hadronization corrections to
  five-jet observables using the event generator SHERPA, which employs
  the CKKW procedure to combine a reliable perturbative treatment of
  high-multiplicity jet final states with parton showers.  We show
  that a competitive value of the strong coupling constant
  $\alpha_s(M_Z) = 0.1156^{+0.0041}_{-0.0034}$ can be extracted from
  the distribution of the five-jet resolution parameter and the
  five-jet rate at LEP1 and LEP2.  }


\begin{document}

\section{Introduction} 

The production of hadrons in $e^+e^-$ annihilation is one of the
best-studied processes in high-energy physics. It played a crucial
role in establishing the correctness of QCD as the theory of strong
interactions. Studies of hadronic final states in $e^+e^-$
annihilation at PEP, KEK, PETRA, SLD and LEP were instrumental for
understanding jets, advancing perturbative QCD computations,
developing parton showers and investigating non-perturbative QCD
effects in high-energy collisions.

At LEP experiments, exclusive processes with up to five jets in the
final state were studied in detail; inclusive measurements are
available for up to six jets~\cite{Buskulic:1996tt,Acton:1993zh,
  Alexander:1996kh,Ackerstaff:1997kk,Abbiendi:1999sx,Abbiendi:2004qz,
  Abbiendi:2008zz,Acciarri:1995ia,Acciarri:1997xr,Acciarri:1998gz,
  Achard:2002kv,Achard:2004sv,Abreu:1999rc,Abdallah:2003xz,
  Abdallah:2004xe,Heister:2003aj}. Theoretically, $e^+e^-$
annihilation cross sections into
two~\cite{Anastasiou:2004qd,Weinzierl:2006ij} and
three~\cite{GehrmannDeRidder:2007hr,GehrmannDeRidder:2008ug,Weinzierl:2008iv}
jets are known at next-to-next-to-leading order (NNLO) in perturbative
QCD, the production of four jets is known at next-to-leading order
(NLO)~\cite{Signer:1996bf,Dixon:1997th,Nagy:1997yn,Nagy:1998bb,Campbell:1998nn}, while only leading order (LO)
predictions were available so far for five-jet observables.  Resummed
results for jet rates have been also
obtained~\cite{Brown:1990nm,Catani:1991hj,Catani:1991pm,
  Dissertori:1995qx,Banfi:2001bz}.

High-quality data, primarily from LEP measurements at the $Z$-pole,
open up an opportunity to perform very accurate studies in jet
physics~\cite{Dasgupta:2003iq}.  Since, by now, gross features of QCD
are well understood, the interest shifts towards subtle details which
can be revealed only through dealing with complicated final states and
improving the accuracy of theoretical predictions.  Because the
$n$-jet rate is proportional to the strong coupling constant at high
power $\sigma_{n {\rm jet}} \sim \alpha_s^{n-2}$, leading-order
predictions for $n$-jet observables for $n \gg 2$ are very uncertain
and an improved theoretical description of such final states is
desirable.  When improved descriptions become available, they are used
to study interesting properties of hadronic final states including
event rates, event shapes and ultimately to extract the value of the
strong coupling constant $\alpha_s$~\cite{Weinzierl:2006ij,
  GehrmannDeRidder:2007hr,GehrmannDeRidder:2008ug,GehrmannDeRidder:2007bj,
  Dissertori:2007xa,Dissertori:2009qa,Weinzierl:2009ms,Dixon:1997th,
  Becher:2008cf,Chien:2010kc}.

Infra-red safe observables, traditionally studied in $e^+e^-$
annihilation, are dominated by short-distance physics; they are
computable in perturbative QCD up to corrections suppressed by inverse
powers of some large energy scale related to e.g.~the average relative
transverse momentum of jets in a given process.  Those power (or
hadronization) corrections are typically estimated using event
generators such as PYTHIA~\cite{Sjostrand:2006za},
HERWIG~\cite{Corcella:2000bw}, and ARIADNE~\cite{Lonnblad:1992tz},
under the assumption of a complete factorization of non-perturbative
and perturbative physics. This factorization implies that the
hadronization corrections to an infrared-safe observable ${\cal O}$
can be estimated as
\be H^{i}[{\cal O}] = \frac{{\cal O}^{i}_{\rm
    hadr}}{{\cal O}^{i}_{\rm part}},
\label{eq1}
\ee
where ${\cal O}^{i}_{\rm hadr}$ and ${\cal O}^{i}_{\rm part}$ are the
values of the observable ${\cal O}$ computed at the hadron and at the
parton level with the event generator $i$.  Because hadronization
corrections are assumed to be factorizable, they can be used to
``improve'' the perturbative prediction ${\cal O}_{\rm pt}$ for the
observable under study.  Hence, one defines the quantity \mbox{${\cal
    O}_{\rm impr} = H^{i}[{\cal O}]\; {\cal O}_{\rm pt}$}, and
compares it to experimental data\footnote{ There are alternative
  approaches to estimate hadronization corrections that address both
  theoretical~\cite{Dokshitzer:1995qm} and
  experimental~\cite{Heister:2003aj,Barate:1996fi} aspects of this
  procedure.}.

While the use of the procedure that we just described is widespread,
it is clear that it can not be fully valid.  Indeed, one can imagine
that, for a particular observable ${\cal O}$, the event generator $i$
happens to reproduce its measured value, \mbox{${\cal O}^{i}_{\rm
    hadr} = {\cal O}_{\rm Data}$}.  As a result, ${\cal O}_{\rm impr}$
reads
\be
{\cal O}_{\rm impr} = 
{\cal O}_{\rm Data}\; \frac{{\cal O}_{\rm pt}}{{\cal O}_{\rm part}^{i}}.
\label{Oimproved}
\ee
Clearly, ${\cal O}_{\rm impr}$ can only be equal to ${\cal O}_{\rm
  data}$ if ${\cal O}_{\rm pt} = {\cal O}_{\rm part}^{i}$, but this
equality can not hold true for a variety of reasons, including
different approximations in parton/dipole showers and fixed-order
perturbative calculations, different dependencies of ${\cal O}_{\rm
  pt}$ and ${\cal O}_{\rm part}^{i}$ on the renormalization scale
$\mu$, etc. Differences at the perturbative level are particularly
worrisome when considering high-multiplicity final states, since
standard event generators routinely used in $e^+e^-$ studies are based
solely on $e^+e^-\to q\bar{q}$ and $e^+e^-\to q\bar{q}g$ matrix
elements. This implies that, since event generators are tuned to data,
hadronization corrections as defined by eq.~(\ref{eq1}) contain {\it
  both} non-perturbative and perturbative effects; the latter are
present to compensate for deficiencies of the partonic part of a
particular event generator\footnote{See ref.~\cite{Dissertori:2009ik}
for a related discussion.}.  
To the extent that both perturbative and
hadronization corrections are small, the inconsistency of the whole
procedure may not be very apparent, but it becomes evident if those
corrections are large.  It is very likely that these issues are
important for five-jet production at LEP. Indeed, because $e^+e^- \to
5~{\rm jets}$ involves a high power of $\alpha_s$, NLO QCD corrections
are expected to be large.  In addition, a correct description of five
hard, well-separated partons is difficult for traditional event
generators, so that all the problems of a conventional approach to
estimating hadronization corrections can be exposed by studying 
five-jet observables.

By extracting hadronization corrections with traditional event
generators -- PYTHIA, HERWIG and ARIADNE -- we find these corrections
to be large and generator-dependent. This is unfortunate since it
implies a large spread of ``improved'' predictions for five-jet
observables when perturbative and hadronization effects are
combined. For this reason, we believe it is important to obtain
hadronization corrections from an event generator whose perturbative
part is up to the task of describing the production of five hard,
well-separated partons.  Within the context of event generation, the
implementation of such a description requires a consistent matching
between high-multiplicity matrix elements, and parton/dipole
showers. In this paper, we employ the SHERPA event generator
\cite{Gleisberg:2003xi,Gleisberg:2008ta}, which implements the
CKKW~\cite{Catani:2001cc} matching prescription\footnote{We emphasize
  that it is not the parton/dipole shower {\em per se} that makes the
  difference, but the possibility to match it to high-multiplicity
  matrix elements. We expect to obtain quantitatively similar results
  with HERWIG, PYTHIA, and ARIADNE if these event generators are
  supplemented with a matrix-element-matching procedure.}.  
By calculating hadronization corrections with SHERPA/CKKW, we find 
that they are relatively small (see the right pane of fig.~\ref{fig1a}), 
in particular in the kinematic region where perturbative QCD 
is reliable. This is what we expect since, with CKKW
matching, the perturbative description of five-jet production provides
a good approximation to the actual physical process. As a consequence,
when we use SHERPA/CKKW to extract non-perturbative corrections
according to eq.~(\ref{eq1}), the results are less contaminated by
perturbative contributions, compared to the case when traditional
event generators are used for this purpose.  Since these
considerations apply to hadron collisions as well, our results have
obvious implications for jet physics at the LHC.

The goal of this paper is to investigate five-jet production at LEP,
using theoretical predictions that are accurate at NLO in QCD.  We
discuss hadronization corrections and show that they depend
significantly upon the event generator that is used to estimate them.
We extract the strong coupling constant by fitting the NLO QCD
predictions for the distributions of the five-jet resolution parameter
$y_{45}$ and the five-jet rate $R_5$. The remainder of the paper is
organized as follows. In Section~\ref{s2} we discuss the technical
details pertinent to the computation of the NLO QCD corrections to
$e^+e^- \to 5~{\rm jets}$.  In Section~\ref{s3} a phenomenological
analysis of five-jet production at LEP is reported.  In
Section~\ref{s4} the value of the strong coupling constant is
extracted. In Section~\ref{s5} we present our conclusions. Details of
the fit procedure that we use in our analysis of the strong coupling
constant are described in the Appendix.

\section{Technical details}
\label{s2}
The computation of NLO QCD corrections to any process or observable in
the context of a subtraction formalism involves the evaluation of the
following contributions: the one-loop virtual corrections, the
real-emission corrections and their subtraction terms, and the finite
remainders obtained from the analytical integration of the subtraction
terms over the degrees of freedom of the unresolved parton(s). The
one-loop virtual amplitudes for $e^+e^- \to 5~{\rm jet}$ required in
this paper are computed within the generalized $D$-dimensional
unitarity framework, as described in
Refs.~\cite{Ellis:2007br,Giele:2008ve,Ellis:2008qc}; a few
technicalities relevant to the five-jet case are given in
sect.~\ref{sec:Rocket}.  The remaining contributions are calculated
using MadFKS~\cite{Frederix:2009yq}, which is also employed to perform
the integration over the phase space of the final short-distance cross
sections.  Details of MadFKS relevant for this computation are
reviewed in sect.~\ref{sec:MadFKS}.

\subsection{Calculation of one-loop amplitudes\label{sec:Rocket}}
Within the context of generalized $D$-dimensional unitarity, we
compute the so-called primitive amplitudes~\cite{Bern:1994fz}, which
are gauge-invariant subsets of color-ordered amplitudes.  The color
decomposition of amplitudes that we need in this paper coincides with
the color decomposition of QCD amplitudes without any colorless vector
boson $\gamma^*/Z$. We use the color decomposition introduced
in ref.~\cite{DelDuca:1999rs}. The relation between primitive and
color-ordered amplitudes that we need in this paper can be found in
ref.~\cite{Ellis:2008qc}.
  
In general, the amplitudes needed for the NLO calculation of
$e^+e^-\to \gamma^*/Z \to 5$ partons are very similar to the
amplitudes with a $W$-boson and five partons.  The main differences
between amplitudes that involve charged and neutral currents originate
from the different couplings of $W$-bosons and $\gamma^*/Z$ bosons to
fermions. In particular, in the Standard Model, the coupling of the
$W$-boson to fermions is only left-handed, while for $\gamma^*/Z$ one
has to sum over left- and right-handed states.  Indeed, the $Z$-boson
couples to fermions via a vector and a vector-axial coupling $g_V
\gamma^\mu + g_A \gamma^\mu\gamma_5$.  We can rewrite this coupling
through left- and right-handed projection operators $P_{\rm L,R} =
(1\mp \gamma_5)/2$ as
\be
g_V \gamma^\mu 
+ g_A \gamma^\mu\gamma_5 = 
  g_L \gamma^\mu P_{\rm L} 
+ g_R \gamma^\mu P_{\rm R}, 
\ee
where $g_{L,R} = g_V \mp g_A$. Therefore, all the one-loop amplitudes
relevant for this paper can be obtained by considering quarks coupled
to a vector current only; the only subtlety is that the vector current
should couple to left- and right-handed quarks with different
strength.

As we already mentioned, many of the non-trivial amplitudes that we
need are identical to the amplitudes calculated for $0\to W+5$ partons
case~\cite{Ellis:2008qc}.  We note, however, that new amplitudes
appear if $\gamma^*/Z$ couples directly to a loop of virtual
fermions. There are two reasons for this. First, such amplitudes are
not present in the $W$-boson case studied in ref.~\cite{Ellis:2008qc}
because of charge (or flavor) conservation.  Second, certain parts of
those amplitudes are related to an axial anomaly and, therefore,
violate the symmetry between the vector current and the axial-vector
current. It is interesting to remark that within the context of
generalized $D$-dimensional unitarity, a correct computation of the
axial anomaly entails a literal implementation of the 't Hooft-Veltman
prescription for dealing with $\gamma_5$ in dimensional
regularization. We have calculated the required amplitudes where
$\gamma^*/Z$ couples to the fermion loop but we neglect them in this
paper since previous experience with those amplitudes shows that they
are very
small~\cite{Kniehl:1989bb,Hagiwara:1990dx,Signer:1996bf,Dixon:1997th},
especially when compared to the residual theoretical uncertainty of
the five-jet rate.
 
The cross section involves Born and virtual amplitudes with two or
four quarks in the final state.  We relate amplitudes involving a
$Z$-boson exchange to amplitudes that only involve the photon (vector)
exchange:
\begin{eqnarray}
&&  A_{Z}(\bar q_\lambda, q_{-\lambda}, e^+_\mu, e^-_{-\mu}, \dots ) 
=  Q_q^{-1} Q_e^{-1} {\cal P}_Z(s) \Bigg [
g_R^q g_R^e\delta_{R\lambda}\delta_{R\mu} A_{\gamma^*}(\bar
q_-, q_{+}, \bar e_-, e_{+}, \dots ) 
\nonumber \\*&&\phantom{aaaaaaa}
+g_R^qg_L^e
\delta_{R\lambda}\delta_{L\mu} A_{\gamma^*}(\bar
q_-, q_{+}, \bar e_+, e_{-}, \dots ) 
+g_L^qg_R^e
\delta_{L\lambda}\delta_{R\mu} A_{\gamma^*}(\bar
q_+, q_{-}, \bar e_-, e_{+}, \dots ) 
\nonumber \\*&&\phantom{aaaaaaa}
+g_L^qg_L^e
\delta_{L\lambda}\delta_{L\mu} A_{\gamma^*}(\bar
q_+, q_{-}, \bar e_+, \bar e_{-}, \dots )\, \Bigg ].  
\label{eq125}
\end{eqnarray}
In eq.~(\ref{eq125}) all particles are outgoing, the dots denote all
gluon momenta, and the propagator factor ${\cal P}(s)$ is given by
\begin{equation}
{\cal P}_Z(s) = \frac{s}{s-M_Z^2+i \Gamma_Z M_Z}\,,
\end{equation}
where $M_Z$ and $\Gamma_Z$ are the mass and the width of the $Z$
boson respectively. The left- and right-handed coupling of
the electrons ($f=e$) and quarks ($f=q$) to the $Z$ boson are
explicitly given by
\begin{equation}
g_L^f = g_V^f-g_A^f = \frac{2(T_3^f- Q^f\sin2\theta_W)}{\sin(2\theta_W)}\,,
\qquad
g_R^f = g_V^f+g_A^f = \frac{-2Q^f\sin2\theta_W}{\sin(2\theta_W)}\,,
\end{equation}
where $\theta_W$ is the Weinberg angle, $T_3^e= T_3^d=-1/2$,
$T_3^u=1/2$ are the values of weak isospin for quarks and leptons and 
$Q^e = -1$, $Q^u= 2/3$, $Q^d=-1/3$ are the respective electric charges.     

The amplitudes involving four quarks and a $Z$-boson can be written in
a similar way. The amplitude that involves different quark
flavors can always be written as the sum of two amplitudes, where
the $Z$-boson couples to a particular quark line
\begin{eqnarray}
A_Z(\bar q_\lambda, q_{-\lambda}, 
\bar Q_{\Lambda}, Q_{-\Lambda},e^+_\mu, e^-_{-\mu}, \dots )
&=& {\tilde A}_Z(\bar q_\lambda, q_{-\lambda}, \bar Q_{\Lambda}, 
Q_{-\Lambda},e^+_\mu, e^-_{-\mu}, \dots )
\nonumber \\*&+&
{\tilde A}_Z( \bar Q_{\Lambda}, Q_{-\Lambda},\bar q_\lambda, 
q_{-\lambda},e^+_\mu, e^-_{-\mu}, \dots ). 
\end{eqnarray}
Note that when amplitudes $\tilde A$ are computed, the $Z$-boson is
allowed to couple to the quark flavor indicated by first and second
argument of $\tilde A$. The expression for $\tilde A_Z$ amplitude in
terms of $\tilde A_\gamma^*$ is identical to eq.~(\ref{eq125}), so we
do not repeat it here.  Finally, if the flavor of the two quark lines
is the same, we include the symmetry factor $1/4$ and anti-symmetrise
with respect to the exchange of the quarks or anti-quarks.  We point
out that the full one-loop matrix elements squared that we use in this
paper were checked against a similar computation performed by the
BlackHat
collaboration~\cite{Berger:2008sz,Berger:2009ep,Berger:2010vm}, and
complete agreement was found.

\subsection{Real emission corrections with MadFKS\label{sec:MadFKS}}
MadFKS is based on the (FKS) subtraction formalism of
ref.~\cite{Frixione:1995ms}.  The implementation of the FKS procedure
is fully automated in MadFKS. In essence, MadFKS goes through the
following steps. First, it determines the partonic processes that
contribute to a given physical reaction, and their singularity
structures. Then, it constructs the real-emission matrix elements,
their subtraction terms, the finite remainders, and the Born matrix
elements.  Finally, it proceeds to the actual computation, by sampling
(possibly with multi-channeling techniques) the phase space, by
evaluating the short-distance cross sections, and by returning
weighted parton-level kinematic configurations, that can be used to
construct as many observables as one likes.  All matrix elements,
except those of virtual origin, are obtained by calls to
MadGraph~\cite{Alwall:2007st} routines. The virtual matrix elements
are on the other hand computed as described in sect.~\ref{sec:Rocket}.
We point out that MadFKS gives, for each phase-space point, a
four-momentum configuration as input to the code of
sect.~\ref{sec:Rocket}, which returns three numbers -- corresponding
to the double and single (IR) pole residues, and to the finite part;
the talk-to between the two codes uses the Binoth-Les~Houches
interface~\cite{Binoth:2010xt}.  As clarified in
ref.~\cite{Frederix:2009yq}, MadFKS integrates quantities that are
locally finite in the phase space, and in four dimensions.  Therefore,
the pole residues provided by the virtual amplitudes are used only to
check that they match those that are known analytically from the
subtraction procedure, in this way ensuring that KLN cancellation does
indeed take place. The only output of the code of
sect.~\ref{sec:Rocket} used in the integration of the short-distance
cross sections is thus the finite part, defined according to the
conventions given in the Appendix~B of ref.~\cite{Frederix:2009yq}.

We have performed the calculation of the five-jet cross section in a
straightforward manner. We included all partonic processes, and
performed explicit sums over colors and helicities, except in the case
of virtual amplitudes. For the latter, we have computed separately the
leading- and subleading-color contributions, and performed the sum
over helicities using Monte Carlo methods. We have used five massless
quark flavors.  We remind the reader that the FKS formalism is
particularly efficient in keeping the number of subtraction terms to a
minimum.  Furthermore, the real-emission matrix element minus the
subtraction terms is re-organized into a sum that gets as many
contributions as the subtraction terms themselves, and that are {\em
  separately} finite, which implies that they can be (and are)
integrated independently from each other. Physically, these
contributions corresponds to pairs of particles (called FKS pairs)
that can give one soft and/or one collinear singularity at most. We
report in table~\ref{tab:FKSpairs} the number of FKS pairs relevant to
the various real-emission processes that contribute to the five-jet
cross section.
%%%%%%%%%%%%%%%%%%%%%%%%%%%%%%%%%%%%%%%%%%%%%%%%%%%%%%%%%%%%%%%%%%%%%
\begin{table}[t]
\begin{center}
\begin{tabular}{lc}\toprule
Process & \# of FKS pairs 
\\\midrule
$\epem\to q\qb gggg$ & 3
\\
$\epem\to q\qb \qp\qpb gg$ & 7
\\
$\epem\to q\qb q\qb gg$ & 4
\\
$\epem\to q\qb \qp\qpb \qpp\qppb$ & 3
\\
$\epem\to q\qb q\qb \qp\qpb$ & 2
\\
$\epem\to q\qb q\qb q\qb$ & 1
\\\bottomrule
\end{tabular}
\end{center}
\caption[FKSpairs] {\label{tab:FKSpairs}{
Numbers of FKS pairs for the various real-emission processes
that contribute to the five-jet cross section. See the text
for details.
}}
\end{table}
%%%%%%%%%%%%%%%%%%%%%%%%%%%%%%%%%%%%%%%%%%%%%%%%%%%%%%%%%%%%%%%%%%%%%

With five massless flavors, the number of independent partonic
subprocesses that contribute to the five-jet cross section is 25.
Using the entries of table~\ref{tab:FKSpairs}, this implies 81 FKS
pairs in total, i.e.~81 independent integrations.  On the other hand,
the complexity of the kinematics is such that even in the context of
an adaptive integration it may be very difficult to map correctly all
the peaks of the Feynman diagrams, and thus to have a stable numerical
behavior. We have therefore preferred to adopt a multi-channelling
integration strategy, that in MadFKS follows the same procedure as in
MadGraph~\cite{Maltoni:2002qb}. In doing so, the numbers of
integration channels we deal with at the real-emission and virtual
level are equal to 3620 and $2\times 1408$ respectively (the factor of
two in the virtual amplitudes being due to the independent integration
of the leading- and subleading-color contributions).  These numbers
are much larger than the 81 FKS pairs we started with; however, the
Feynman-diagrams peaks can now be mapped accurately by the integration
routines, and relatively small statistics is sufficient in each
channel to obtain numerical stability.  We conclude by stressing that
the MadFKS integration channels are fully independent. Furthermore,
they are not determined dynamically (e.g.~by performing a preliminary
integration of the cross section), but are defined {\em a priori}, by
considering the topologies of the Feynman diagrams that contribute to
the relevant partonic processes. The whole organization of the
calculation is therefore inherently parallel.

\section{Phenomenology of five-jet production}
\label{s3}

In this Section we present the results of our calculation.  We
consider $e^+e^- \to {\rm jets}$ and define jets using the Durham jet
algorithm~\cite{Catani:1991hj} with resolution parameter $y_{\rm
  cut}$.  The following distance between each pair of particles is
used in the Durham jet algorithm
\be
y_{ij} = \frac{2 {\rm min} (E_i^2, E_j^2)}{s} 
\left (1 - \cos \theta_{ij} \right ), 
\label{Durham}
\ee
where $s$ is the center-of-mass energy of the collision squared,
$E_{i}$ is the energy of the parton $i$, and $\theta_{ij}$ is the
relative angle between the partons $i$ and $j$, in the $e^+e^-$
center-of-mass reference frame. The pair of particles with the
smallest distance is clustered together by adding their four-momenta,
as long as $y_{ij} < y_{\rm cut}$, and the procedure is then iterated.
When all distances $y_{ij}$ are larger than $y_{\rm cut}$, the
recombination stops and the number of jets in the event is defined to
be equal to the number of (pseudo)-particles left at that stage.

In this paper we consider two observables which we define with the
Durham jet algorithm.  The first observable is the differential
distribution with respect to the five-jet resolution parameter
$y_{45}$, normalized to the total cross section for $e^+e^- \to {\rm
  hadrons}$, $\sigma_{\rm tot}$ (which we compute at the NLO,
i.e.~at ${\cal O}(\as)$).  The resolution parameter $y_{45}$ is
the maximal value of $y_{\rm cut}$ such that a given event is
classified as a five-jet event by the Durham jet algorithm. We note
that \be \frac{1}{\sigma_{\rm tot}} \int \limits_{y_{\rm cut}}^{1}
{\rm d}y_{45} \; \frac{{\rm d} \sigma}{ {\rm d}y_{45}} =
\frac{\sigma_{\rm incl}^{5-{\rm jet}}(y_{\rm cut})}{\sigma_{\rm tot}},
\ee where $\sigma_{\rm incl}^{5-{\rm jet}}$ is the {\it inclusive}
five-jet production cross section in $e^+e^-$ annihilation.  The
second observable that we study is the five-jet rate $R_5(y_{\rm
  cut})$. It is defined as follows
\be
R_5(y_{\rm cut}) = 
\frac{\sigma_{\rm excl}^{5-{\rm jet}}(y_{\rm cut})}{\sigma_{\rm tot}},
\ee
where $\sigma_{\rm excl}^{5-{\rm jet}}(y_{\rm cut})$ is the exclusive
five-jet production cross section. It is calculated by applying the
Durham jet algorithm to the given event, and by requiring that exactly
five jets are reconstructed.

When we compute $\sigma^{-1}_{\rm tot} {\rm d}\sigma/{\rm d } \ln
y_{45}^{-1}$ and $R_5$ in perturbative QCD, we obtain a power series
in the strong coupling constant
\ba
\sigma^{-1}_{\rm tot} \frac{{\rm d}\sigma}{{\rm d } \ln y_{45}^{-1}}
&=& \left ( \frac{\alpha_s(\mu)}{2\pi} \right )^3 A_{45}(y_{45}) 
+ 
\left ( \frac{\alpha_s(\mu)}{2\pi} \right )^4 
\left ( B_{45}(y_{45}) 
+ 3 b_0 A_{45}(y_{45}) \ln \frac{\mu}{\sqrt{s}} \right ),\;
\label{y45sig}
\\
R_5(y_{\rm cut}) &=& 
\left ( \frac{\alpha_s(\mu)}{2\pi} \right )^3 A_{5}(y_{\rm cut}) 
+ 
\left ( \frac{\alpha_s(\mu)}{2\pi} \right )^4 
\left ( B_{5}(y_{\rm cut}) 
+ 3 b_0 A_{5}(y_{\rm cut}) \ln \frac{\mu}{\sqrt{s}} \right ),\;
\label{R5sig}
\ea
where $\mu$ is the renormalization scale, $b_0 = (33 - 2n_f)/3$ and
$n_f = 5$ is the number of quark flavors that we treat as massless.
The top quark is considered to be infinitely heavy and is completely
neglected in our computation. It is important to emphasize that the
coefficients $A_{45,5}$ and $B_{45,5}$ depend on $y_{45}$ and $y_{\rm
  cut}$, respectively, but not, say, on the total center-of-mass
energy squared. This feature is a consequence of the following
approximations employed in our computation:
1) all particles, except the $Z$ boson, are treated as {\it massless};
2) the observables that we are interested in are sufficiently {\it
  inclusive} so that the vector and the axial currents do not
interfere;
3) we neglect triangle fermion diagrams that lead to the axial anomaly
so that (for equal couplings) vector and axial current contributions
to the final result are equal.
These three points are sufficient to ensure that $A_{45,5}$ and
$B_{45,5}$ are independent of the electroweak parameters and the 
center-of-mass energy squared. 

%%%%%%%%%%%%%%%%%%%%%%%%%%%%%%%%%%%%%%%%%%%%%%%%%%%%%%%%%%%%%%%%%%%%%
\begin{figure}[t!]
\begin{center}
\includegraphics[angle=0,scale=0.5]{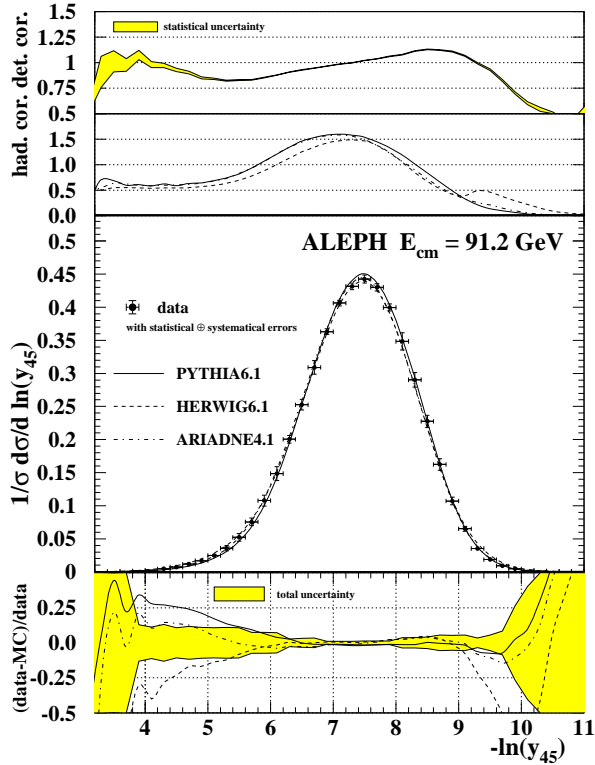}
\caption{ALEPH data~\cite{Heister:2003aj} for the $y_{45}$ 
distribution at LEP1, compared to PYTHIA, HERWIG and 
ARIADNE results. The upper panes show  
detector and hadronization corrections, respectively.
The lowest pane shows the relative difference between 
data and event generator predictions. 
This figure was provided to us by H.~Stenzel.
}
\label{fig1}
\end{center}
\end{figure}
%%%%%%%%%%%%%%%%%%%%%%%%%%%%%%%%%%%%%%%%%%%%%%%%%%%%%%%%%%%%%%%%%%%%%

Experimentally~\cite{Heister:2003aj}, five-jet observables are
computed using the reconstructed momenta and energies of charged and
neutral particles.  Measurements are corrected for detector effects,
so that final distributions correspond to stable hadrons and leptons,
and for initial- and final-state photon radiation, which is a sizable
correction for LEP2 data. Above the $Z$ peak, relevant backgrounds are
subtracted; the most important among them is $W$-pair production. 
The experimental uncertainties are estimated by varying event- and
particle-selection cuts.  They are below 1\% at LEP1 and slightly
larger at LEP2.  Further details of the experimental analysis can be
found in ref.~\cite{Heister:2003aj}.

In fig.~\ref{fig1}, we compare ALEPH LEP1 data~\cite{Heister:2003aj}
for $1/{\sigma_{\rm tot}}{\rm d}\sigma/{\rm d} \ln y_{45}^{-1}$ with
the hadron-level predictions of three event generators -- PYTHIA,
HERWIG and ARIADNE.  We observe that these event generators describe
experimental data fairly well; differences between data and
theoretical predictions are below twenty five percent in the central $
4.5 < \ln y_{45}^{-1} < 9 $ region of the distribution, where the
statistical accuracy of the data is good.  This is an impressive
accomplishment since $\sigma_{\rm tot}^{-1} {\rm d} \sigma/{\rm d} \ln
y_{45}^{-1}$ changes by three orders of magnitude in this range of
$\ln y_{45}^{-1}$.  On the other hand, it is clear from the upper pane
of fig.~\ref{fig1} that hadronization corrections are very large and
change from 0.5 to 1.5 in that range of $\ln y_{45}^{-1}$. In
addition, it follows from fig.~\ref{fig1} that the difference between
hadronization corrections, as calculated using different event
generators, can be as large as 20-30$\%$.

\begin{figure}[t!]
\begin{center}
\includegraphics[angle=-90,scale=0.27]{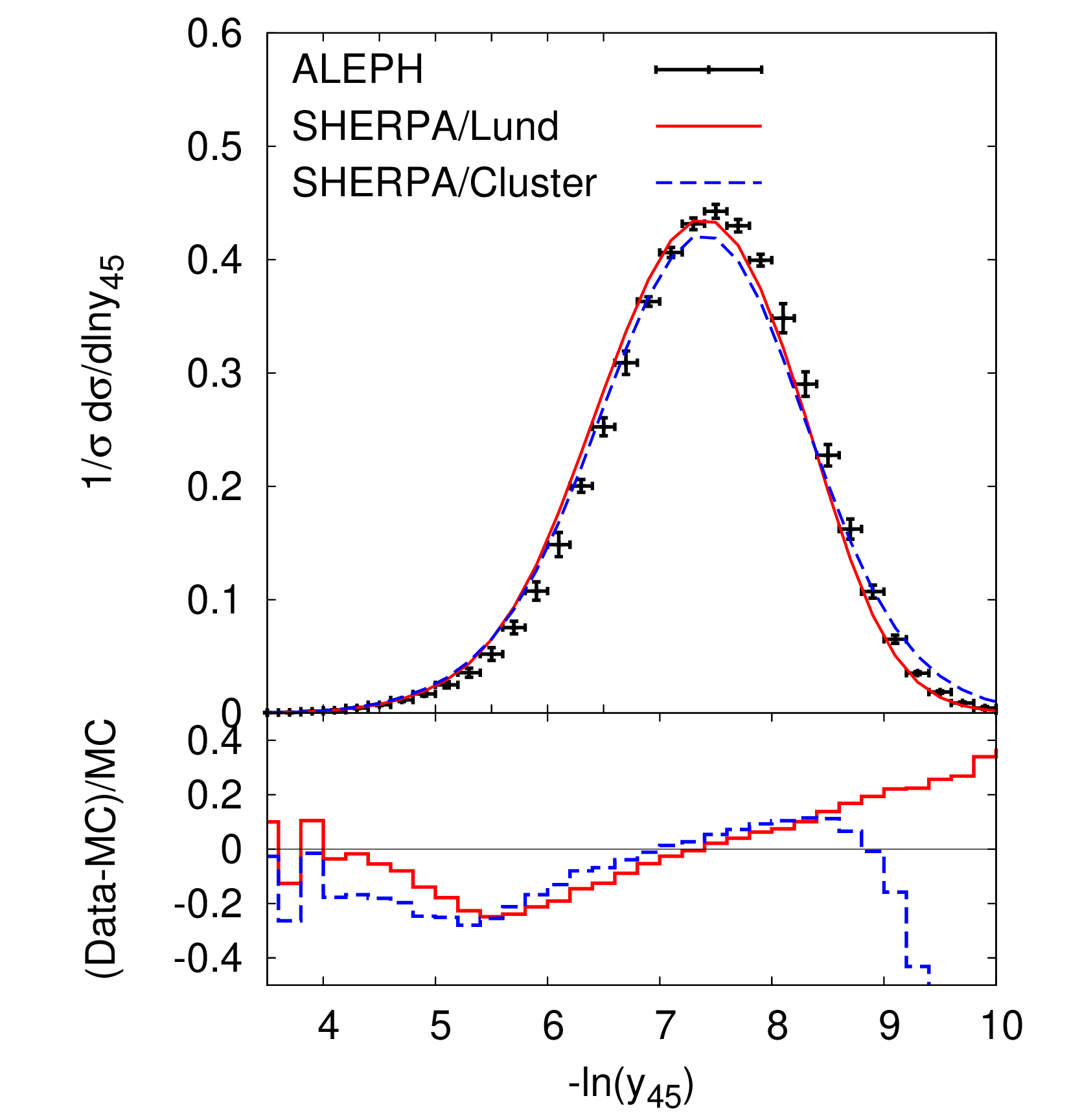}
\includegraphics[angle=-90,scale=0.26]{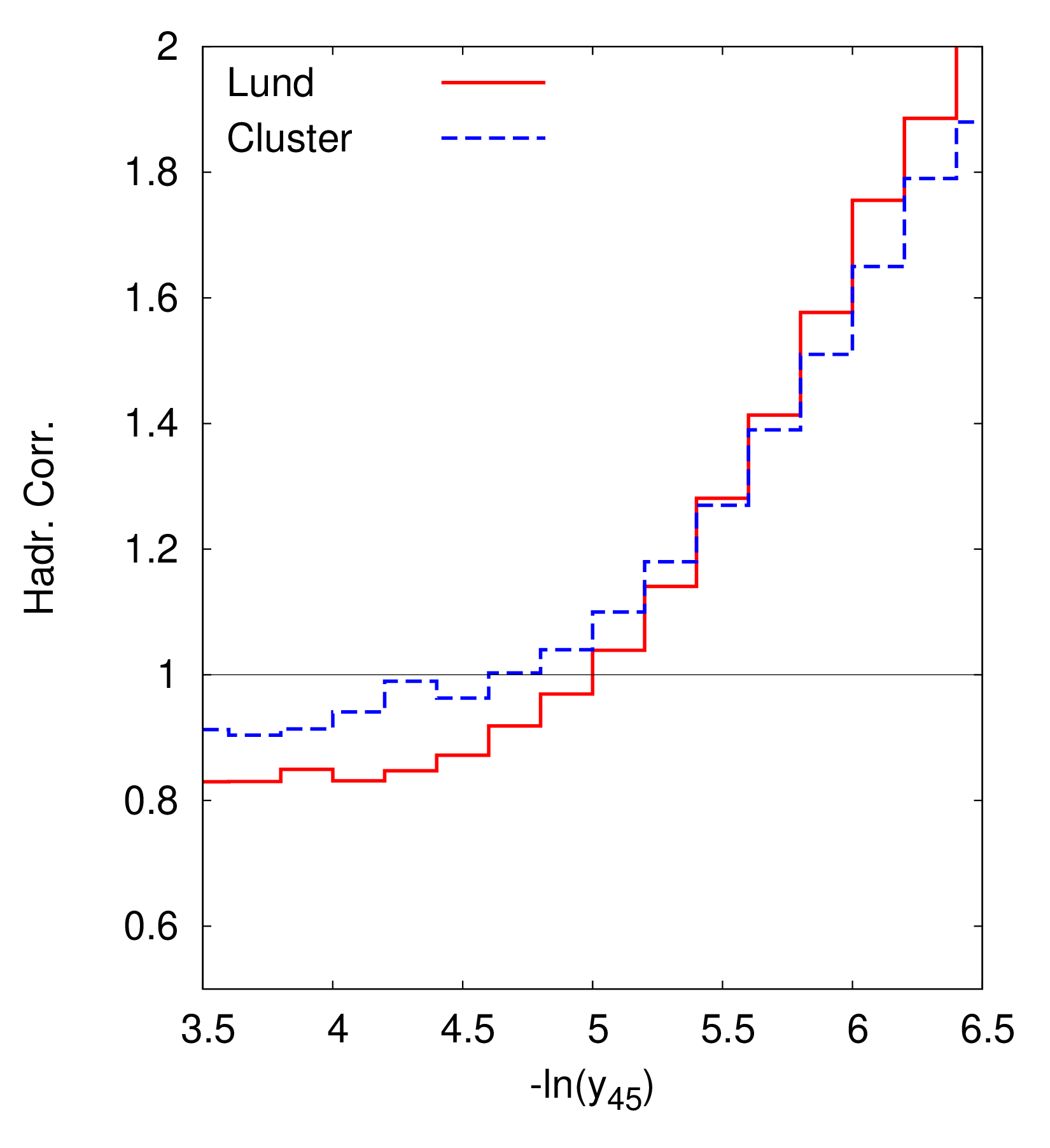}
\caption{ALEPH data for the $y_{45}$ distribution at LEP1, compared to
  SHERPA results. Two hadronization models -- Lund
  string~\cite{Andersson:1983ia} and cluster~\cite{Winter:2003tt} --
  are employed.  The lower pane in the left plot shows the
  relative difference between Sherpa predictions with the two hadronization
  models, and ALEPH data. In the right plot, the hadronization
  corrections for the two models are shown.}
\label{fig1a}
\end{center}
\end{figure}

We attribute these features to the inability of PYTHIA, HERWIG and
ARIADNE to describe hard perturbative radiation correctly. Indeed,
these programs generate high-multiplicity final states starting from
hard {\it low-multiplicity} processes; they produce additional jets by
means of parton/dipole showers. Since these showers describe hard
large-angle emissions only approximately, the so-called hadronization
corrections attempt to correct for this (perturbative) deficiency.
While this problem is unavoidable if traditional event generators are
used to describe high-multiplicity final states, techniques exist to
match parton showers and high-multiplicity matrix elements in a
consistent manner, thereby improving the pure-perturbative part of
event generators.  One such technique is the CKKW matching
procedure~\cite{Catani:2001cc}, which is implemented as default in the
SHERPA event generator.  The comparison of ALEPH LEP1 data with SHERPA
predictions, as well as the hadronization corrections derived from
SHERPA, are shown in fig.~\ref{fig1a}.  Two hadronization models --
Lund string~\cite{Andersson:1983ia} and cluster~\cite{Winter:2003tt}
-- are employed.  In the central part of the distribution, SHERPA
results agree with ALEPH data to $20-25\%$, similar to traditional
event generators. Moreover, in the region of moderately small values
of $\ln y_{45}^{-1}$, where fixed-order perturbative description is
reliable, the hadronization corrections are below twenty percent, {\it in
  sharp contrast} with estimates of hadronization corrections based on
PYTHIA, HERWIG and ARIADNE.  It is important to emphasize that,
although in that region of $\ln y_{45}^{-1}$ traditional event
generators provide slightly better description of data compared to  SHERPA,
this does not mean that hadronization corrections extracted with the
former codes are more reliable.  Indeed, traditional event generators
achieve agreement with data at the price of very large hadronization
corrections. This feature precludes a clear separation between long-
and short-distance phenomena, which is crucial for the procedure
outlined below eq.~(\ref{eq1}) to be meaningful.

%%%%%%%%%%%%%%%%%%%%%%%%%%%%%%%%%%%%%%%%%%%%%%%%%%%%%%%%%%%%%%%%%%%%%
%%%%%%%%%%%%%%%%%%%%%%%%%%%%%%%%%%%%%%%%%%%%%%%%%%%%%%%%%%%%%%%%%%%%%

The ALEPH data exhibit a characteristic turnover shape. This turnover
means that for small values of $y_{45}$, the result is dominated by
exclusive five-jet production with very small resolution parameter,
where fixed order perturbation theory fails and a resummation is
required to achieve meaningful results.  A resummation of $\alpha_s^n
L^{2n}$ and $\alpha_s^n L^{2n-1}$ terms, where $L=\ln y_{\rm
  cut}^{-1}$, was performed for $R_5$ in~\cite{Catani:1991hj}, while
no resummation is currently available for the five-jet resolution
parameter distribution.  However, there seems to be no region in $L$
where this resummation can be valid since two conditions $L\gg 1$ and
$\alpha_s L \ll 1$ should be satisfied simultaneously.  Taking
$\alpha_s \sim 0.15$ as a typical value of the strong coupling
constant\footnote{ We take $5 - 20~{\rm GeV}$ as a reasonable estimate
  of the scale of the strong coupling constant for five-jet production
  in the range $ 3 < \ln y_{\rm cut}^{-1} < 7$.}, we find that $L$
should be {\it smaller} than $6$.  On the other hand, practical
experience with resummations suggests that $L \gg 5$ is what can be
considered as a large logarithm.  Clearly, $5 \ll L < 6$ leaves very
little room for the validity of this approach.  It should be possible
to improve on the resummation by including sub-leading logarithms and
matching to NLO QCD computations.  However, since we do not perform
any resummation in this paper, we require $\ln y_{45}^{-1} ,\ln y_{\rm
  cut}^{-1} \lsim 6$ for the comparison of the NLO QCD computation
with data.  Interestingly, a similar upper bound on $\ln y_{45}^{-1}$
appears because we neglect the mass of $b$-quarks in our
computation. This implies that the resolution parameter times the
center of mass energy should be larger than the $b$-quark mass,
i.e.~$sy_{45} > m_b^2$, which translates into $\ln(y_{45}^{-1}) < \ln
(s/m_b^2) \lsim 6$, for $s = M_Z^2$.

%%%%%%%%%%%%%%%%%%%%%%%%%%%%%%%%%%%%%%%%%%%%%%%%%%%%%%%%%%%%%%%%%%%%%
\begin{figure}[t!]
\begin{center}
\includegraphics[angle=-90,scale=0.26]{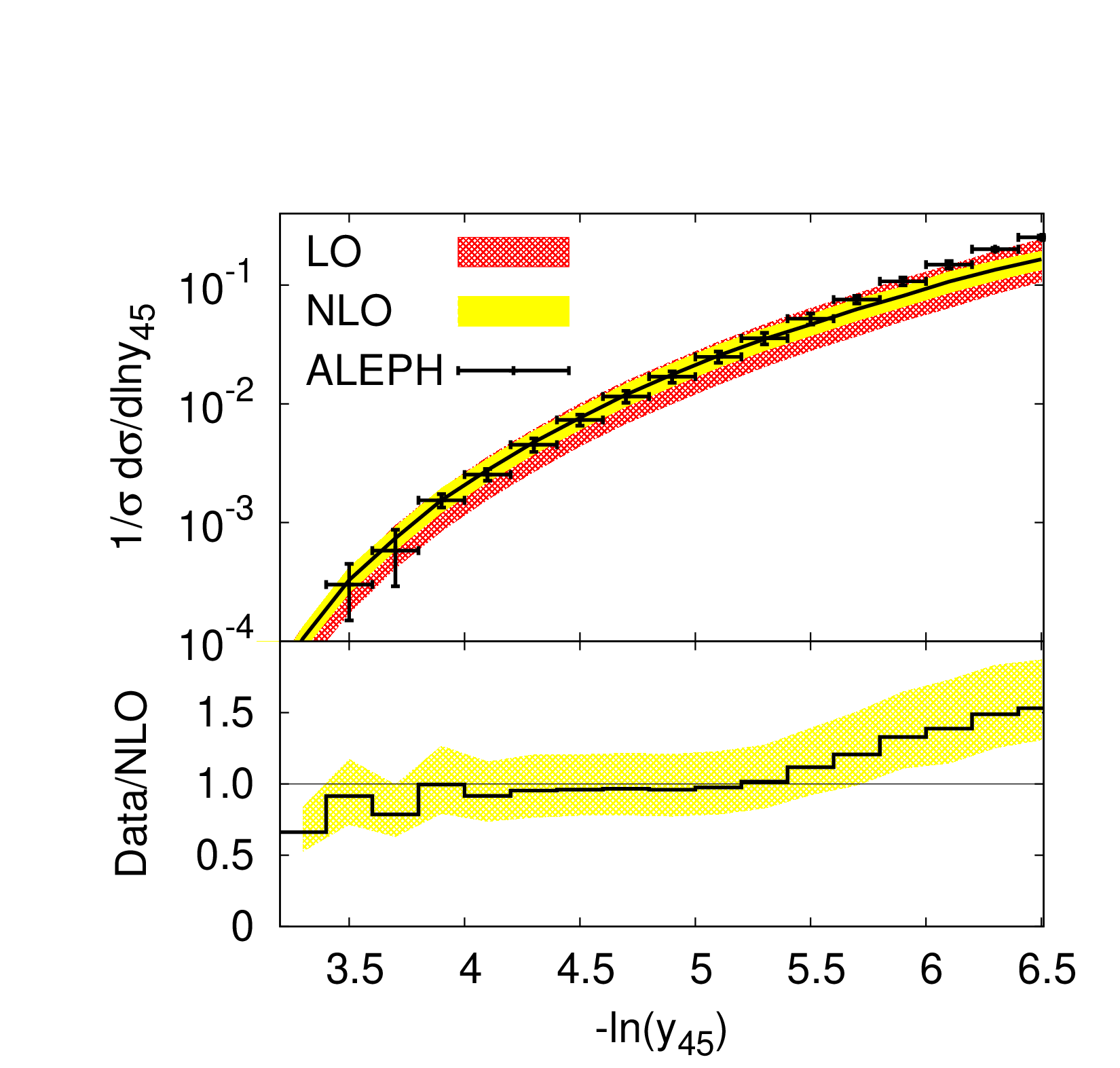}
\includegraphics[angle=-90,scale=0.26]{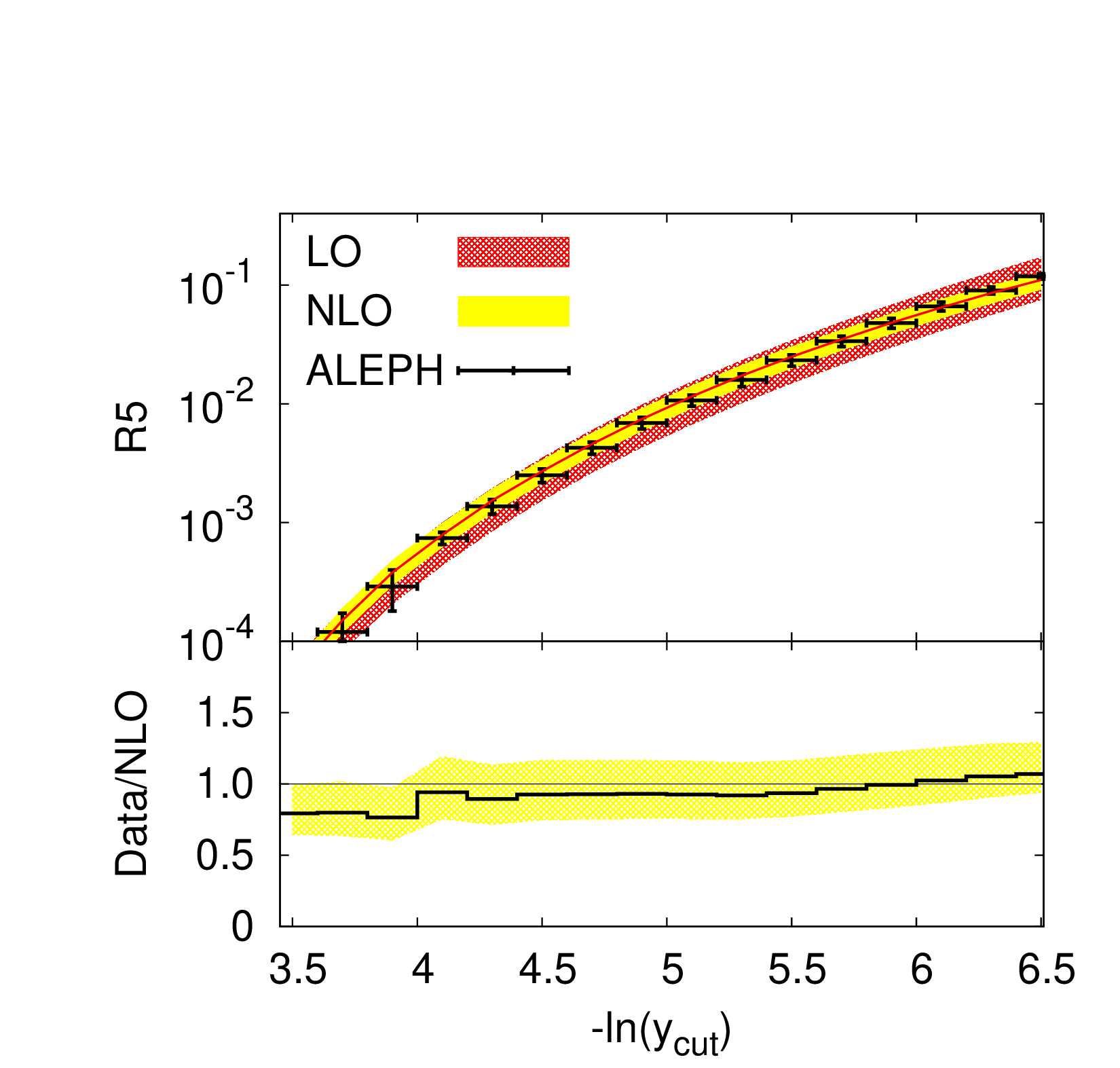}
\caption{ALEPH LEP1 data compared to leading and next-to-leading order
  predictions in QCD, without hadronization corrections.  We use
  $\alpha_s(M_Z) = 0.130$ at the leading and $\alpha_s(M_Z) = 0.118$ at
  the next-to-leading order in perturbative QCD. The renormalization scale
  is chosen to be $0.3M_Z$. The uncertainty bands are obtained by
  considering the scale variation $ 0.15~M_Z < \mu < 0.6~M_Z$.  Solid
  lines refer to NLO QCD results evaluated with $\mu = 0.3 M_Z$.  }
\label{fig2}
\end{center}
\end{figure}
%%%%%%%%%%%%%%%%%%%%%%%%%%%%%%%%%%%%%%%%%%%%%%%%%%%%%%%%%%%%%%%%%%%%

When fixed-order perturbative QCD calculations are compared to
experimental data, the choice of the renormalization scale becomes an
important issue. Traditionally, multi-jet observables in $e^+e^-$
annihilations are computed in perturbative QCD by evaluating the
strong coupling constant at the center-of-mass energy.  However, for
large numbers of jets this choice should be reconsidered, since the
hardness of each jet decreases with their number. Dynamical
renormalization scales used in event generators account for this
effect by relating the choice of the renormalization scale to the
event kinematics. Our choice of the renormalization scale is also
motivated by dynamical considerations. To this end, we consider the
clustering history of five- and six-parton configurations that results
from using the Durham jet algorithm.  We compute the average value of
$\sqrt{y_{23}}$, where $y_{23}$ is the three-jet resolution parameter,
using only phase-space weights. We find this average to be
approximately equal to $0.3$.  Since $\sqrt{y_{23} s}$ is, roughly,
the relative transverse momentum of the hardest branching in the
clustering history, we select $\mu = 0.3\sqrt{s}$ as the default
choice for the renormalization scale of $\alpha_s$ that we use to
describe the five-jet production.

With this choice of the renormalization scale, we compare in
fig.~\ref{fig2} ALEPH LEP1 data for $\sigma_{\rm had}^{-1} {\rm
  d}\sigma_{45}/{\rm d} \ln y_{45}^{-1}$ and $R_5$ with our leading
and next-to-leading results.  We use the value of strong coupling
constant $\alpha_s(M_Z) = 0.130$ for leading order computations and
$\alpha_s(M_Z) = 0.118$ for next-to-leading order computations.  While
it is not customary to change the value of the strong coupling
constant from one order in perturbation theory to the other in
applications of QCD to $e^+e^-$ physics, it is done routinely in the
context of hadron collider physics.  Our choice of the leading-order
value for $\alpha_s$ is motivated by fits of parton distribution
functions and of the strong coupling constant described in
ref.~\cite{Martin:2009iq}; our next-to-leading order value for
$\alpha_s$ is close to the world average~\cite{pdg,Bethke:2009jm}.  We
do not apply hadronization corrections at this stage.  In order to
assess the perturbative uncertainty, a scale variation by a factor of
two around the default scale $\mu = 0.3~M_Z$ is performed.  A close
inspection of the two plots shown in fig.~\ref{fig2} reveals that the
most important effect of the NLO QCD corrections is the reduction in
the uncertainty related to the renormalization scale dependence.  The
renormalization scale uncertainty is reduced from $[-30\%,+45\%]$ at
leading order to $[-20\%,+25\%]$ at next-to-leading order. When
leading order and next-to-leading order results are compared for $\mu
= 0.3~M_Z$, the QCD corrections increase the leading order predictions
by $10-20\%$\footnote{We note that had we used the same value of the
  coupling constant at LO as at NLO, as is usually done in $e^+e^-$
  calculations, NLO corrections would have been much larger,
  $45-60\%$.}.  The agreement between NLO QCD results and ALEPH data
is very good for both observables considered.  However, for the
$\sigma_{\rm had}^{-1} {\rm d}\sigma_{45}/{\rm d} \ln y_{45}^{-1}$
distribution systematic differences appear for $\ln y_{45}^{-1} >
5.2$, whereas the $R_5$ data can be described by fixed-order QCD
without hadronization corrections all the way up to $\ln y_{\rm
  cut}^{-1} = 6.5$.

%%%%%%%%%%%%%%%%%%%%%%%%%%%%%%%%%%%%%%%%%%%%%%%%%%%%%%%%%%%%%%%%%%%%%%%%%%
\begin{table}[t]
\begin{center}
\begin{tabular}{ccc}\toprule
 & LEP1, hadr. & LEP1, no hadr. \\ 
 & $\sigma_{\rm tot}^{-1}{\rm d}\sigma/{\rm d}y_{45}$, $R_5$ & 
$\sigma_{\rm tot}^{-1}{\rm d}\sigma/{\rm d}y_{45}$, $R_5$ \\ 
\midrule\vspace{3pt}
 stat.  & 
$\begin{array}{c} 
+0.0002 \\
-0.0002
\end{array} 
$ 
& 
$\begin{array}{c} 
+0.0002 \\
-0.0002
\end{array} 
$ 
\\\vspace{3pt}
 syst.  & 
$\begin{array}{c} +0.0027 \\
-0.0029
\end{array}                        
 $ 
& 
$\begin{array}{c} +0.0027 \\
-0.0029
\end{array}                        
 $ 
\\\vspace{3pt}
 pert.  &  
$\begin{array}{c} +0.0062 \\
-0.0043
\end{array}
$                        
& 
$\begin{array}{c} +0.0068 \\
-0.0047
\end{array}
$
\\\vspace{3pt}
 fit~range  &  
$\begin{array}{c} +0.0014 \\
-0.0014
\end{array}
$                        
& 
$\begin{array}{c} +0.0005 \\
-0.0005
\end{array}
$\\
 hadr.  &  
$\begin{array}{c} +0.0012 \\
-0.0012
\end{array}
$                        
& -- \\\midrule
$ \alpha_s(M_Z)$ & 
$0.1159  
\begin{array}{c} {+0.0070} \\
{-0.0055}
\end{array}
$
& 
$0.1163  
\begin{array}{c} {+0.0073} \\
{-0.0055}
\end{array}
$
\\ \bottomrule
\end{tabular}
\end{center} 
\caption{\label{tab:fitresultLEP1}{Values of the 
    strong coupling constant $\as(M_Z)$ obtained from fits 
    to ALEPH LEP1 data for $\sigma_{\rm tot}^{-1}{\rm d}\sigma/{\rm d}y_{45}$ 
    and $R_5$. NLO QCD predictions are used. Hadronization  
    corrections are estimated with  SHERPA.  Default fit ranges
    are \mbox{$3.8\le -\ln y_{45}\le 5.2$}, and
    \mbox{$4.0\le -\ln y_{\rm cut}\le 5.6$}.
    See the text for details.
  }}
\end{table}
%%%%%%%%%%%%%%%%%%%%%%%%%%%%%%%%%%%%%%%%%%%%%%%%%%%%%%%%%%%%%%%%%%%%%%%%%%
%%%%%%%%%%%%%%%%%%%%%%%%%%%%%%%%%%%%%%%%%%%%%%%%%%%%%%%%%%%%%%%%%%%%%%%%%%
\begin{table}[t]
\begin{center}
\begin{tabular}{cccc}\toprule
 & LEP2, no hadr. & LEP2, no hadr. & LEP2, no hadr.  \\ 
 & $\sigma_{\rm tot}^{-1}{\rm d}\sigma/{\rm d}y_{45}$ & $R_5$ & 
$\sigma_{\rm tot}^{-1}{\rm d}\sigma/{\rm d}y_{45}$, 
$R_5$  \\ \midrule\vspace{3pt}
 stat.  & 
$\begin{array}{c} + 0.0020 \\
-0.0022
\end{array} 
$ 
& 
$\begin{array}{c} + 0.0022 \\
-0.0025
\end{array} 
$ 
& 
$\begin{array}{c} 
+ 0.0015 \\
-0.0016
\end{array} 
$ 
\\\vspace{3pt}
 syst.  & 
$\begin{array}{c} 
+0.0008 \\
-0.0009
\end{array}                        
 $ 
& 
$\begin{array}{c} +0.0012 \\
-0.0012
\end{array}                        
 $ 
& 
$\begin{array}{c} 
+0.0008 \\
-0.0008
\end{array}                        
 $ 
\\\vspace{3pt}
 pert.  &  
$\begin{array}{c} 
+0.0049 \\
-0.0034
\end{array}
$
& 
$\begin{array}{c} +0.0029 \\
-0.0020
\end{array}
$
& 
$\begin{array}{c} 
+0.0029 \\
-0.0020
\end{array}
$
\\
 fit~range  &  
$\begin{array}{c} 
+0.0038 \\
-0.0038
\end{array}
$
& 
$\begin{array}{c} +0.0030 \\
-0.0030
\end{array}
$
& 
$\begin{array}{c} 
+0.0028 \\
-0.0028
\end{array}
$
\\\midrule
$ \alpha_s(M_Z)$ & 
$0.1189  
\begin{array}{c} {+0.0066} \\
{-0.0057}
\end{array}
$
& 
$0.1120 
\begin{array}{c} {+0.0050} \\
{-0.0047}
\end{array}
$
& 
$0.1155  
\begin{array}{c} {+0.0044} \\
{-0.0039}
\end{array}
$
\\ \bottomrule
\end{tabular}
\end{center} 
\caption{\label{tab:fitresultLEP2}{Values of the 
    strong coupling constant $\as(M_Z)$ obtained from fits 
    to ALEPH LEP2 data with 
    $ E_{\rm cm} \ge 183~{\rm GeV}$ 
    for $\sigma_{\rm tot}^{-1}{\rm d}\sigma/{\rm d}y_{45}$ 
    and $R_5$. NLO QCD predictions are used. 
    Hadronization corrections are not included.
    Default fit ranges
    are \mbox{$4.8\le -\ln y_{45}\le 6.4$}, and
    \mbox{$2.1\le -\log_{10} y_{\rm cut}\le 2.9$}.
    See the text for details.
  }}
\end{table}
%%%%%%%%%%%%%%%%%%%%%%%%%%%%%%%%%%%%%%%%%%%%%%%%%%%%%%%%%%%%%%%%%%%%%%%%%%

\section{The strong coupling constant from  five-jet observables}
\label{s4}
It follows from the discussion in the previous Section that the use of
the world-average value of the strong coupling constant results in
good agreement between parton-level NLO predictions, and LEP1 data
for the five-jet resolution parameter and the five-jet rate.
Therefore, we can turn this consideration around, and extract the
value of the strong coupling constant from these two five-jet
observables at LEP1 and LEP2. To combine the values of the strong
coupling constant extracted from different observables and at
different energies, we use the procedure advocated by the LEP QCD
working group~\cite{Jones:2006nq}.  Technical details of the fit
procedure are reported in the Appendix.

We consider ALEPH data~\cite{Heister:2003aj} for $R_5$ and
$\sigma_{\rm tot}^{-1} {\rm d} \sigma/{\rm d} y_{45}$, as measured at
LEP1 \mbox{($\sqrt{s}=M_Z$)}, and at LEP2 ($\sqrt{s}=183$, 189, 200, and
206~GeV). We point out that data for $\sigma_{\rm tot}^{-1} {\rm d}
\sigma/{\rm d} y_{45}$ are not available at $\sqrt{s}=200$~GeV.
Similar to other precision-QCD studies at LEP, we do not consider LEP2
data taken at $\sqrt{s}<183$~GeV.  As explained above, we use SHERPA
for the computation of hadronization corrections. By default,
hadronization in SHERPA is performed with the {\em cluster}
model~\cite{Winter:2003tt}, which we therefore adopt as our default as
well. Incidentally, we observe that it gives smaller hadronization
corrections in the kinematic range of interest, than the Lund string
hadronization model \cite{Andersson:1983ia}, which is also available
in SHERPA. We use the Lund string model to estimate systematic
uncertainties related to hadronization effects.

Since we use fixed-order perturbative results and do not perform any
resummation, it is not possible to describe the data in the full
kinematic ranges studied by experiments. This feature makes the choice
of the kinematic range used in the fit an important but,
unfortunately, somewhat a subjective issue.  In general, we attempt to
take the fit range as large as possible, with the condition that our
computations are reliable and that the data quality is good. In the
determination of the central value of $\alpha_s$ at LEP1, we consider
\mbox{$3.8\le -\ln y_{45}\le 5.2$} (7 data points) for the five-jet
resolution parameter distribution, and \mbox{$4.0\le -\ln y_{\rm
    cut}\le 5.6$} (8 data points) for $R_5$.  In order to estimate the
error on $\alpha_s$ related to our choice of the fit range, we extract
the value of $\alpha_s$ by performing a second fit, with larger ranges
\mbox{$3.4\le -\ln y_{45}\le 5.6$} (11 data points) for the five-jet
resolution parameter, and \mbox{$3.4\le -\ln y_{\rm cut}\le 6.0$} for
$R_5$ (13 data points).  The difference between the values of $\as$
obtained in the two fits is called the ``fit range'' error; it is
supposed to quantify the uncertainty on $\as$ due to the choice of the
data points included in the fits.

At LEP2 the situation is different. Firstly, data are given with a
coarser binning and, secondly, large fluctuations are present in
experimental results at small values of $\ln y_{45}^{-1}$ and $\ln
y_{\rm cut}^{-1}$ (for example, for some center-of-mass energies the
corresponding observables are not even monotonic).  Because of this,
we decided to exclude those data points from our fits, effectively
reducing the fit ranges.  We note that those data would have had a
modest impact on the final result anyhow, because they are affected by
fairly large errors.  We use \mbox{$4.8\le -\ln y_{45}\le 6.4$} for
the five-jet resolution parameter (2 data points per $\sqrt{s}$), and
\mbox{$2.1\le -\log_{10} y_{\rm cut}\le 2.9$} for $R_5$ (4 data points
per $\sqrt{s}$), to find the central values of $\as$.  In order to
estimate the fit-range error, we employ \mbox{$4.8\le -\ln y_{45}\le
  5.6$} (1 data point per $\sqrt{s}$), and \mbox{$2.1\le -\log_{10}
  y_{\rm cut}\le 2.5$} (2 data points per $\sqrt{s}$), since the
choice of these ranges leads to the {\it largest} changes in the values
of the strong coupling constant compared to the $\as$ values obtained
from fitting with the default ranges.

The results of our fits to LEP1 data are shown in
table~\ref{tab:fitresultLEP1}. The agreement between the two values of
$\as(M_Z)$ extracted with and without hadronization corrections is
impressive; the difference is completely negligible compared to the
overall uncertainties. This result could have been anticipated by
inspecting fig.~\ref{fig1a}, which shows that, in the fit region,
hadronization corrections are small, in particular when the default
SHERPA choice, the cluster model, is used.  We note that if we use
the hadronization corrections as given by conventional HERWIG, PYTHIA,
or ARIADNE, without matching them to high-multiplicity matrix
elements, the picture changes drastically and the values of $\as(M_Z)$
extracted with or without hadronization corrections are quite
different from each other.  We also note that the overall errors of
the two results given in table~\ref{tab:fitresultLEP1} are slightly
smaller when including hadronization corrections. This is due to a
marginally better description of the data in the central region of the
fit range -- which leads to smaller value of $\alpha_s(M_Z)$ and
thus to smaller perturbative errors. However, the error reduction is
partially compensated by the degradation of the fit quality when
including larger values of $\ln y_{45}^{-1}, \ln y_{\rm cut}^{-1}$,
where hadronization corrections increase. This feature leads to larger
fit-range error compared to the no-hadronization case.  Note also that
if we extract the values of $\as(M_Z)$ by
fitting\footnote{Hadronization corrections are included.}  the
five-jet resolution parameter distribution and $R_5$ separately, we
obtain $0.1168^{+0.0076}_{-0.0060}$ and $0.1151^{+0.0071}_{-0.0056}$
respectively.  These values are consistent with the result of the
combined fit shown in table~\ref{tab:fitresultLEP1} but have slightly
larger errors.  From table~\ref{tab:fitresultLEP1}, it is clear that
the sensitivity of the five-jet observables to $\as$ is very high, as
illustrated by the tiny statistical errors. This sensitivity is
ultimately related to the high power of $\as$ that enters the five-jet
observables. In spite of this, the overall error is not particularly
small, since the perturbative uncertainty is still quite sizable at
this order in perturbative QCD.

Compared to LEP1, there are important differences when we extract
$\as$ by fitting to the LEP2 data. Firstly, because hadronization
corrections are negligible at LEP1, and because these corrections
decrease with energy, we do not consider them for LEP2.  Secondly, for
the reasons explained above, we do not consider the data points at
small values of $\ln y_{45}^{-1}$ and $\ln y_{\rm cut}^{-1}$.  This
fact, combined with coarser binning of data, pushes us to the region
of $y_{45}$ that may be affected by large logarithms of the resolution
parameter. As a result, we find larger fit-range errors at LEP2 than
at LEP1.  The statistical errors are also much larger at LEP2 than at
LEP1, as one expects given the luminosities collected.  On the other
hand, since the effective strong coupling is smaller at LEP2, the
perturbative uncertainty affecting five-jet observables decreases,
making the $\as$ extraction at LEP2 competitive with that done at
LEP1.
This effect is particularly strong for $R_5$, for which the extracted
central value of $\as$ is slightly smaller than for the five-jet
resolution parameter.
 In table~\ref{tab:fitresultLEP2} we present the
$\as$ values obtained by fitting separately the five-jet resolution
parameter and $R_5$ at LEP2, since they differ from each other by a
larger amount than at LEP1. Still, both values are within one
standard deviation from the strong coupling constant that we obtain by
performing a simultaneous fit to the two observables.  We take the latter 
value, given in the third column of table~\ref{tab:fitresultLEP2}, as our
best determination of $\as$ from LEP2 data.

We obtain our final estimate of the strong coupling constant by
combining the values of $\as(M_Z)$ extracted from LEP1 and LEP2 data.
We assume that the statistical and systematic errors of the two
results are not correlated (an assumption which is strictly correct
for the former, and a very good approximation for the latter), while
the perturbative errors are considered to be fully correlated.  The
correlation of the perturbative uncertainties is due to the fact that
we estimated them by varying the renormalization scale, which results
in changes of the cross sections whose pattern is independent of the
center-of-mass energy. It is quite likely that a more sophisticated
approach to estimating perturbative errors (see
e.g.~ref.~\cite{Jones:2003yv}) will result in a smaller uncertainty on
$\alpha_s$.  Hence, the procedure that we employ in this paper is
rather conservative.  Using the results of
tables~\ref{tab:fitresultLEP1} and~\ref{tab:fitresultLEP2}, we finally
obtain 
\be 
\alpha_s(M_Z) = 0.1156^{+0.0041}_{-0.0034}\,.
\label{eqsherpa1}
\ee
We note that if we perform the fit to {\it both} LEP1 and LEP2 
data simultaneously, we obtain 
\mbox{$\alpha_s(M_Z)=0.1156^{+0.0045}_{-0.0041}$},
in perfect agreement with eq.~(\ref{eqsherpa1}).

The value of $\alpha_s(M_Z)$ that we extract from five-jet observables
at LEP can be compared with other recent determinations of this
quantity, shown in table~\ref{tab:tab2}. We see that both the central
value of $\alpha_s$ and its error, obtained from fitting five-jet
observables, compare well with other determinations.  On the other
hand, it is interesting that $\alpha_S(M_Z)$ in eq.~(\ref{eqsherpa1})
is lower than the world average.  It is peculiar that a number of
recent determinations of $\alpha_s$ arrived at a similar conclusion.
%%%%%%%%%%%%%%%%%%%%%%%%%%%%%%%%%%%%%%%%%%%%%%%%%%%%%%%%%%%%%%%%%%%%%%%%%%
\begin{table}[t]
\begin{center}
\begin{small}
\begin{tabular}{ccc}\toprule
Observable & $\alpha_s(M_Z)$ & Ref. \\ \midrule
 $\tau$ decays  & $0.1197 \pm 0.0016 $ & \cite{Bethke:2009jm} \\
 $\Upsilon$ decays   & $0.119~\, \pm 0.0055 $ & \cite{Brambilla:2007cz} \\
 3~jet observables  & $0.1224 \pm 0.0039$ & \cite{Dissertori:2009ik} \\
 jets in DIS  & $0.1198 \pm 0.0032$ & \cite{Glasman:2007sm} \\
 DIS  &   $0.1142 \pm 0.0021 $ & \cite{Blumlein:2007dk} \\
 thrust &      $0.1135 \pm 0.0011$   & \cite{Abbate:2010xh} \\ 
 lattice &  $0.1183 \pm 0.0008 $ & \cite{Davies:2008sw}\\
 EW fits &  $ 0.1193 \pm 0.0028$ & \cite{Flacher:2008zq}\\
\midrule
 world average &      $0.1184 \pm 0.0007$   & \cite{Bethke:2009jm} \\ 
\midrule
 $e^+e^-\to$~five~jets &      $0.1156 \pm 0.0038$   & this paper \\ 
\bottomrule
\end{tabular}
\end{small}
\end{center} 
\caption{\label{tab:tab2}{Summary of selected determinations 
of the  strong coupling constant $\alpha_s(M_Z)$. We have averaged the 
$\pm$ errors shown eq.~(\protect\ref{eqsherpa1}) for the determination 
of $\alpha_s$ reported in this paper. Not all results shown in this table 
are included in the world
average.
}}
\end{table}
%%%%%%%%%%%%%%%%%%%%%%%%%%%%%%%%%%%%%%%%%%%%%%%%%%%%%%%%%%%%%%%%%%%%%%%%%%

\section{Conclusions}
\label{s5}
In this paper we study the production of five jets in $e^+e^-$
annihilation at LEP1 and LEP2.  We improve the perturbative QCD
predictions for five-jet observables, $1/\sigma_{\rm tot}{\rm
  d}\sigma/{\rm d}y_{45}$ and $R_5$, by computing the NLO QCD
corrections.  For suitably chosen renormalization scales, such
corrections are between ten and twenty percent\footnote{Note that this
  statement is only valid if the leading order result is calculated
  with $\alpha_S(M_Z) = 0.130$, which is much larger than the value of
  the strong coupling constant given in eq.~(\ref{eqsherpa1}) or the
  world average. Changing the value of the strong coupling
  constant from one order in perturbation theory to the other is not
  customary in $e^+e^-$ collider physics, while it is an accepted
  practice in hadron collider physics.}.  They reduce the scale
uncertainty by about a factor of two with respect to the LO
predictions, and lead to a better agreement between theoretical
predictions and experimental data.

We point out that hadronization corrections computed with event
generators whose showers are not matched to high-multiplicity matrix
elements (such as out-of-the-box HERWIG, PYTHIA, and ARIADNE) are
large and uncertain.  For this reason, we believe it is important to
describe five-jet observables in a way that incorporates
high-multiplicity tree-level matrix elements.  This is provided by the
event generator SHERPA, which implements the CKKW procedure for
matching tree-level matrix elements to parton showers. In this way, an
improved description of five hard, well-separated partons is obtained,
which in turn results in fairly small hadronization corrections in the
range where fixed-order perturbative results are most reliable.

We extract the strong coupling constant from the distributions of the
five-jet resolution parameter and the five-jet rate, as measured at
LEP1 and LEP2 by the ALEPH collaboration. We find \mbox{$\alpha_s(M_Z)
  = 0.1156^{+0.0041}_{-0.0034}$}, which compares well with other
recent determinations of the strong coupling constant, and is somewhat
lower than the current world average value.  We stress that our
treatment of the uncertainties on $\as$ is conservative. A detailed
knowledge of the experimental systematics, and a more sophisticated
approach to theoretical errors will very likely lead to a higher
precision in the determination of $\alpha_s$ from five-jet observables
at LEP.

\section*{Acknowledgments} 
We are grateful to Hasko~Stenzel for the participation in earlier
stages of this work.  We would like to thank Stefan H\"oche for
providing us with SHERPA results.  Useful conversations with
Andrea~Banfi, Guenther~Dissertori, Stefano~Forte, Michelangelo~Mangano, 
Gavin~Salam, Peter~Skands, Roberto~Tenchini, Paolo~Torrielli, and
Bryan~Webber are gratefully acknowledged.  During the work on this
paper, we have benefited from the hospitality extended to us by the
Aspen Center for Physics, the Fermilab Theory group, and the Theory
Unit at CERN.  This research is supported by the NSF under grant
PHY-0855365, by the start-up funds provided by Johns Hopkins
University, by the British Science and Technology Facilities Council,
and by the Swiss National Science Foundation under contract
200020-126691.

\section*{Appendix: details of the fit} 
In this Appendix, the details of the fitting procedure are described.
In the fit, we consider two observables -- the five-jet resolution
parameter distribution and the five-jet rate. These observables are
measured at several energies; the results are available in the form of
binned distributions.  In principle, we can extract the value of the
strong coupling constant from {\it any} of the bins but, clearly, the
availability of many bins helps in decreasing the errors. The problem
is that both the experimental and theoretical errors affecting
different bins may be correlated, and it is important to take these
correlations correctly into account, to obtain a proper estimate of
the overall uncertainty in the determination of $\as$.  Our procedure
follows closely the approach of the LEP QCD working group described in
ref.~\cite{Jones:2006nq}.  A possible way to treat theoretical
uncertainties is discussed in ref.~\cite{Jones:2003yv}, but we follow
a simpler approach that is described below.

The set of all available bins, for a chosen range of $y_{45}$ and
$y_{\rm cut}$, is a set of observables $L_{\cal O}$ from which values
of the strong coupling constant can be determined.  Each member of
$L_{\cal O}$ is represented by three numbers \mbox{${\cal O}_i =
  [X_i,\sigma_i^{\rm stat},\sigma_i^{\rm syst}]$}, where $X_i$, and
$\sigma_i^{\rm stat,syst}$ are the central value and the statistical
and systematic uncertainties respectively, for a given observable in
the bin $i$.  If uncertainties are asymmetric, ${\cal O}_i$ is a
collection of five numbers, and the discussion below applies to
positive and negative errors separately. In what follows, we use
$\as\pm\delta\as$ as the shorthand notation instead of the full one
${\as}^{+\delta^{\scriptscriptstyle +}\!\!\as}_
{-\delta^{\scriptscriptstyle -}\!\!\as}$.

As we already mentioned, each of the observables from the list
$L_{\cal O}$ can be used to determine the value of the strong coupling
constant. This is done by solving the equation
\be
T_i \;H_i = E_i,
\label{eq_a_1}
\ee
where $T_i$ is the (parton level) theoretical prediction, $H_i$ is the
hadronization correction, and $E_i$ is the experimental value for the
bin $i$.  The theoretical prediction $T_i$ depends on $\alpha_s(M_Z)$
{\it and} the renormalization scale $\mu$.  As discussed in the text,
for the hadronization correction we can use either the cluster or the
Lund string model and we choose the former as our default.  We write
the value of the strong coupling constant at $M_Z$, obtained by
solving eq.~(\ref{eq_a_1}), as
\be
\alpha_s^{i} = {\overline \alpha}_s^{i} 
\pm \delta \alpha_s^{i,\rm stat}
\pm \delta \alpha_s^{i,\rm syst}
\pm \delta \alpha_s^{i,\rm scale}
\pm \delta \alpha_s^{i,\rm hadr},
\label{eq_a_2}
\ee
where ${\overline \alpha}_s^{i}$ is the central value. The central
value and the errors in eq.~(\ref{eq_a_2}) are obtained in the
following way:
\begin{itemize} 

\item[{\em 1)}]
the central value ${\overline \alpha}_s^{i}$ is obtained by solving 
eq.~(\ref{eq_a_1}) for $\alpha_s$ with $\mu = \mu_0$, 
$\mu_0 = 0.3\sqrt{s}$, $E_i = X_i$ (i.e., the central data value in
the relevant bin), and the cluster model for hadronization. The 
results without hadronization corrections are obtained by simply 
setting $H_i=1$.

\item[{\em 2)}]
$\delta\alpha_s^{i,\rm stat}$ and $\delta\alpha_s^{i,\rm syst}$
are obtained by solving eq.~(\ref{eq_a_1}) for $\as$ with the
same settings as in item {\em 1)}, except that
$E_i = X_i\pm\sigma^{i,\rm stat}$ or $E_i = X_i\pm\sigma^{i,\rm syst}$.
The differences between the values of $\as$ obtained in this way,
and the central value  ${\overline \alpha}_s^{i}$, are 
$\pm\delta\alpha_s^{i,\rm stat}$ and $\pm\delta\alpha_s^{i,\rm syst}$.

\item[{\em 3)}]
$\delta\alpha_s^{i,\rm scale}$ is obtained by solving eq.~(\ref{eq_a_1}) 
for $\as$ with the same settings as in item {\em 1)}, except 
that $\mu=0.5\mu_0$ and $\mu=2\mu_0$ are  used.
The differences between the values of $\as$ obtained in this way,
and the central value  ${\overline \alpha}_s^{i}$, are 
$\pm\delta\alpha_s^{i,\rm scale}$.

\item[{\em 4)}]
$\delta\alpha_s^{i,\rm hadr}$ is obtained by solving eq.~(\ref{eq_a_1}) 
for $\as$ with the same settings as in item {\em 1)}, 
except that 
the Lund string model is used for hadronization. We define 
$\pm\delta\alpha_s^{i,\rm hadr}=\pm |\as-{\overline \alpha}_s^{i}|$.
We note that this is a conservative choice, since clearly
$\as-{\overline \alpha}_s^{i}$ is either positive or negative.

\end{itemize}
The result of this procedure is a set of values of the strong coupling
constants, $\alpha_s^{i}$, with the corresponding errors.  They need
to be combined to obtain the average value of $\alpha_s$.  To this
end, it is necessary to construct the covariance matrix, which we
define as the sum of the covariance matrices for statistical,
systematic, perturbative, and hadronization errors.  If we denote
generically by $\dasi$ one of these four errors, the corresponding
covariance matrix is
\be
V_{ij}=\delta_{ij}\left(\dasi\right)^2
+\left(1-\delta_{ij}\right)\,C_{ij}\dasi\dasj\,,
\label{Vclassic}
\ee
where $C_{ij}$ is the statistical correlation between $\alpha_s^i$ and
$\alpha_s^j$ (see later).  The covariance matrix is used to calculate
the average value of the strong coupling constant and its error in a
standard way.  We compute the weights
\be
w_i = \sum_{j=1}^{N} (V^{-1})_{ij}\Big/\sum_{k,l=1}^{N}(V^{-1})_{kl},
\;\;\;\; 1 \le i \le N\,,\;\;\; N = {\rm dim}(V),
\label{omegadef}
\ee
and obtain the estimate of the average of the strong coupling constant and 
of its error
\ba
&& \alpha_s = \sum_{i=1}^{N} w_i {\bar \alpha}_s^{i}\,,
\label{asomega}
\\
&& \sigma^2(\alpha_s) = \sum_{i,j=1}^{N} w_i V_{ij} w_j\,.
\label{sigomega}
\ea
It is easy to see that if two or more errors that enter the definition
of the covariance matrix are close numerically, and the absolute
values of the corresponding statistical correlations $C_{ij}$ is close
to one, eq.~(\ref{Vclassic}) may lead to pathological results, since
the weights $w_i$ tend to grow large in absolute value and to have
opposite signs.  To avoid this, the LEP QCD working
group~\cite{Jones:2006nq} adopts the formula
\be
V_{ij}=\delta_{ij}\left(\dasi\right)^2
+\left(1-\delta_{ij}\right)\,\min\Big\{(\dasi)^2,(\dasj)^2\Big\}\,,
\label{VLEPQCD}
\ee
which is essentially equivalent to taking the largest possible
$C_{ij}$ in eq.~(\ref{Vclassic}), that still leads to non-pathological
weights $w_i$.  Clearly, eq.~(\ref{VLEPQCD}) is an overestimate of the
correlation if the actual $C_{ij}$ is small.  For this reason, in our
fit we use a slightly modified formula for the covariance matrix
\be
V_{ij}=\delta_{ij}\left(\dasi\right)^2
+\left(1-\delta_{ij}\right)\,
\min\Big\{(\dasi)^2,(\dasj)^2,C_{ij}\dasi\dasj\Big\}\,.
\label{Vours}
\ee
As we explain below, for the observables and errors that we consider,
eq.~(\ref{Vours}) coincides with eq.~(\ref{VLEPQCD}) in all cases,
except for the statistical correlation between $\sigma_{\rm
  tot}^{-1}{\rm d}\sigma/{\rm d}y_{45}$ and $R_5$.

In summary, we take the  covariance matrix to be
\be
V=V^{\rm stat}+V^{\rm syst}+V^{\rm scale}+V^{\rm hadr}\,,
\label{fullV}
\ee
where each of the terms on the right-hand side of eq.~(\ref{fullV}) is
constructed according to eq.~(\ref{Vours}) using
$\delta\alpha_s^{i,\rm stat}$, $\delta\alpha_s^{i,\rm syst}$,
$\delta\alpha_s^{i,\rm scale}$, and $\delta\alpha_s^{i,\rm hadr}$
respectively. As far as the off-diagonal terms of the various $V$
matrices are concerned, we have assumed what follows:
\begin{itemize}
\item Statistical errors are uncorrelated between different
  center-of-mass energies.  At a given center-of-mass energy,
  $\sigma_{\rm tot}^{-1}{\rm d}\sigma/{\rm d}y_{45}$ data are
  uncorrelated, $R_5$ data are fully correlated, and $R_5$ data are
  correlated with $\sigma_{\rm tot}^{-1}{\rm d}\sigma/{\rm d}y_{45}$
  ones for $y_{\rm cut}\le y_{45}$. We have explicitly computed the
  coefficient $C_{ij}$ relevant to the $\sigma_{\rm tot}^{-1}{\rm
    d}\sigma/{\rm d}y_{45}-R_5$ correlation, and have used them in
  eq.~(\ref{Vours}).

\item Since we do not have detailed information about correlations of
  systematic uncertainties in ALEPH data, we assume conservatively
  that all systematic errors at a given center-of-mass energy are
  fully correlated.  We also assume that systematic errors are
  completely uncorrelated between LEP1 and LEP2, but that they
  are fully correlated in the measurements performed at different
  LEP2 energies. 

\item Perturbative errors are assumed to be fully correlated, for all
  observables and energies. See also the main text for a comment on
  this point.

\item Hadronization errors, that we compute only at LEP1, are assumed
  to be fully correlated.
\end{itemize}
Finally we point out that in the computation of the central value of
$\as$, according to eq.~(\ref{asomega}), we neglect the off-diagonal
entries of $V^{\rm scale}$ and $V^{\rm hadr}$ \cite{Jones:2006nq}. On
the other hand, we include all the off-diagonal entries in the
computation of the standard deviation according to
eq.~(\ref{sigomega}). In fact, in the presence of errors numerically
very close to each other (which is the case for the perturbative and
hadronization errors), the result for the average value of $\as$ tends
to assume the value of the input with the smallest error, which is
again an artifact of the combination procedure.  We have checked that the
value of the strong coupling constant that we obtain in this way is
statistically fully compatible with the result we would have obtained
by considering all correlations when determining the central value of
$\as$.

\bibliography{fivejet}{}

\providecommand{\href}[2]{#2}\begingroup\raggedright\begin{thebibliography}{10}

\bibitem{Buskulic:1996tt}
{\bf ALEPH} Collaboration, D.~Buskulic {\em et~al.}, {\it {Studies of QCD in
  $e^+ e^- \to$ hadrons at $E_{cm} =$ 130 and 136 GeV}},  {\em Z.Phys.} {\bf
  C73} (1997) 409--420.

\bibitem{Acton:1993zh}
{\bf OPAL} Collaboration, P.~Acton {\em et~al.}, {\it {A Determination of
  $\alpha_s(M_{Z^0})$ at LEP using resummed QCD calculations}},  {\em Z.Phys.}
  {\bf C59} (1993) 1--20.

\bibitem{Alexander:1996kh}
{\bf OPAL} Collaboration, G.~Alexander {\em et~al.}, {\it {QCD studies with
  $e^+ e^-$ annihilation data at 130 and 136 GeV}},  {\em Z.Phys.} {\bf C72}
  (1996) 191--206.

\bibitem{Ackerstaff:1997kk}
{\bf OPAL} Collaboration, K.~Ackerstaff {\em et~al.}, {\it {QCD studies with
  $e^+ e^-$ annihilation data at 161 GeV}},  {\em Z.Phys.} {\bf C75} (1997)
  193--207.

\bibitem{Abbiendi:1999sx}
{\bf OPAL} Collaboration, G.~Abbiendi {\em et~al.}, {\it {QCD studies with $e^+
  e^-$ annihilation data at 172--189 GeV}},  {\em Eur.Phys.J.} {\bf C16} (2000)
  185--210, [\href{http://xxx.lanl.gov/abs/hep-ex/0002012}{{\tt
  hep-ex/0002012}}]. Revised Feb 2000.

\bibitem{Abbiendi:2004qz}
{\bf OPAL} Collaboration, G.~Abbiendi {\em et~al.}, {\it {Measurement of event
  shape distributions and moments in $e^+ e^- \to$ hadrons at 91--209 GeV and a
  determination of $\alpha_s$}},  {\em Eur.Phys.J.} {\bf C40} (2005) 287--316,
  [\href{http://xxx.lanl.gov/abs/hep-ex/0503051}{{\tt hep-ex/0503051}}].

\bibitem{Abbiendi:2008zz}
{\bf OPAL} Collaboration, G.~Abbiendi {\em et~al.}, {\it {Measurement of
  $\alpha_s$ with radiative hadronic events}},  {\em Eur.Phys.J.} {\bf C53}
  (2008) 21--39.

\bibitem{Acciarri:1995ia}
{\bf L3} Collaboration, M.~Acciarri {\em et~al.}, {\it {Study of the structure
  of hadronic events and determination of $\alpha_s$ at $\sqrt{s} =$ 130 and
  136 GeV}},  {\em Phys.Lett.} {\bf B371} (1996) 137--148.

\bibitem{Acciarri:1997xr}
{\bf L3} Collaboration, M.~Acciarri {\em et~al.}, {\it {QCD studies and
  determination of $\alpha_s$ in $e^+ e^-$ collisions at $\sqrt{s} =$ 161 and
  172 GeV}},  {\em Phys.Lett.} {\bf B404} (1997) 390--402.

\bibitem{Acciarri:1998gz}
{\bf L3} Collaboration, M.~Acciarri {\em et~al.}, {\it {QCD results from
  studies of hadronic events produced in $e^+ e^-$ annihilations at $\sqrt{s} =
  183$ GeV}},  {\em Phys.Lett.} {\bf B444} (1998) 569--582.

\bibitem{Achard:2002kv}
{\bf L3} Collaboration, P.~Achard {\em et~al.}, {\it {Determination of
  $\alpha_s$ from hadronic event shapes in $e^+ e^-$ annihilation at 192 GeV
  $\leq \sqrt{s} \leq$ 208 GeV}},  {\em Phys.Lett.} {\bf B536} (2002) 217--228,
  [\href{http://xxx.lanl.gov/abs/hep-ex/0206052}{{\tt hep-ex/0206052}}].

\bibitem{Achard:2004sv}
{\bf L3} Collaboration, P.~Achard {\em et~al.}, {\it {Studies of hadronic event
  structure in $e^+ e^-$ annihilation from 30 to 209 GeV with the L3
  detector}},  {\em Phys.Rept.} {\bf 399} (2004) 71--174,
  [\href{http://xxx.lanl.gov/abs/hep-ex/0406049}{{\tt hep-ex/0406049}}].

\bibitem{Abreu:1999rc}
{\bf DELPHI} Collaboration, P.~Abreu {\em et~al.}, {\it {Energy dependence of
  event shapes and of $\alpha_s$ at LEP2}},  {\em Phys.Lett.} {\bf B456} (1999)
  322--340.

\bibitem{Abdallah:2003xz}
{\bf DELPHI} Collaboration, J.~Abdallah {\em et~al.}, {\it {A Study of the
  energy evolution of event shape distributions and their means with the DELPHI
  detector at LEP}},  {\em Eur.Phys.J.} {\bf C29} (2003) 285--312,
  [\href{http://xxx.lanl.gov/abs/hep-ex/0307048}{{\tt hep-ex/0307048}}].

\bibitem{Abdallah:2004xe}
{\bf DELPHI} Collaboration, J.~Abdallah {\em et~al.}, {\it {The Measurement of
  $\alpha_s$ from event shapes with the DELPHI detector at the highest LEP
  energies}},  {\em Eur.Phys.J.} {\bf C37} (2004) 1--23,
  [\href{http://xxx.lanl.gov/abs/hep-ex/0406011}{{\tt hep-ex/0406011}}].

\bibitem{Heister:2003aj}
{\bf ALEPH} Collaboration, A.~Heister {\em et~al.}, {\it {Studies of QCD at
  $e^+ e^-$ centre-of-mass energies between 91 and 209 GeV}},  {\em
  Eur.Phys.J.} {\bf C35} (2004) 457--486.

\bibitem{Anastasiou:2004qd}
C.~Anastasiou, K.~Melnikov, and F.~Petriello, {\it {Real radiation at NNLO:
  $e^+ e^- \to$ 2 jets through $\mathcal{O}(\alpha_s^2)$}},  {\em
  Phys.Rev.Lett.} {\bf 93} (2004) 032002,
  [\href{http://xxx.lanl.gov/abs/hep-ph/0402280}{{\tt hep-ph/0402280}}].

\bibitem{Weinzierl:2006ij}
S.~Weinzierl, {\it {NNLO corrections to 2-jet observables in electron-positron
  annihilation}},  {\em Phys.Rev.} {\bf D74} (2006) 014020,
  [\href{http://xxx.lanl.gov/abs/hep-ph/0606008}{{\tt hep-ph/0606008}}].

\bibitem{GehrmannDeRidder:2007hr}
A.~Gehrmann-De~Ridder, T.~Gehrmann, E.~Glover, and G.~Heinrich, {\it {NNLO
  corrections to event shapes in $e^+ e^-$ annihilation}},  {\em JHEP} {\bf
  0712} (2007) 094, [\href{http://xxx.lanl.gov/abs/arXiv:0711.4711}{{\tt
  arXiv:0711.4711}}].

\bibitem{GehrmannDeRidder:2008ug}
A.~Gehrmann-De~Ridder, T.~Gehrmann, E.~Glover, and G.~Heinrich, {\it {Jet rates
  in electron-positron annihilation at $\mathcal{O}(\alpha_s^3)$ in QCD}},
  {\em Phys.Rev.Lett.} {\bf 100} (2008) 172001,
  [\href{http://xxx.lanl.gov/abs/arXiv:0802.0813}{{\tt arXiv:0802.0813}}].

\bibitem{Weinzierl:2008iv}
S.~Weinzierl, {\it {NNLO corrections to 3-jet observables in electron-positron
  annihilation}},  {\em Phys.Rev.Lett.} {\bf 101} (2008) 162001,
  [\href{http://xxx.lanl.gov/abs/arXiv:0807.3241}{{\tt arXiv:0807.3241}}].

\bibitem{Signer:1996bf}
A.~Signer and L.~J. Dixon, {\it {Electron--positron annihilation into four jets
  at next-to-leading order in $\alpha_s$}},  {\em Phys.Rev.Lett.} {\bf 78}
  (1997) 811--814, [\href{http://xxx.lanl.gov/abs/hep-ph/9609460}{{\tt
  hep-ph/9609460}}].

\bibitem{Dixon:1997th}
L.~J. Dixon and A.~Signer, {\it {Complete $O(\alpha_s^3)$ results for $e^+
  e^-\to (\gamma, Z) \to$ four jets}},  {\em Phys.Rev.} {\bf D56} (1997)
  4031--4038, [\href{http://xxx.lanl.gov/abs/hep-ph/9706285}{{\tt
  hep-ph/9706285}}].

\bibitem{Nagy:1997yn}
Z.~Nagy and Z.~Trocsanyi, {\it {Next-to-leading order calculation of four jet
  shape variables}},  {\em Phys.Rev.Lett.} {\bf 79} (1997) 3604--3607,
  [\href{http://xxx.lanl.gov/abs/hep-ph/9707309}{{\tt hep-ph/9707309}}].

\bibitem{Nagy:1998bb}
Z.~Nagy and Z.~Trocsanyi, {\it {Next-to-leading order calculation of four jet
  observables in electron positron annihilation}},  {\em Phys.Rev.} {\bf D59}
  (1999) 014020, [\href{http://xxx.lanl.gov/abs/hep-ph/9806317}{{\tt
  hep-ph/9806317}}].

\bibitem{Campbell:1998nn}
J.~M. Campbell, M.~Cullen, and E.~Glover, {\it {Four jet event shapes in
  electron - positron annihilation}},  {\em Eur.Phys.J.} {\bf C9} (1999)
  245--265, [\href{http://xxx.lanl.gov/abs/hep-ph/9809429}{{\tt
  hep-ph/9809429}}].

\bibitem{Brown:1990nm}
N.~Brown and W.~Stirling, {\it {Jet cross-sections at leading double logarithm
  in $e^+ e^-$ annihilation}},  {\em Phys.Lett.} {\bf B252} (1990) 657--662.

\bibitem{Catani:1991hj}
S.~Catani, Y.~L. Dokshitzer, M.~Olsson, G.~Turnock, and B.~Webber, {\it {New
  clustering algorithm for multi-jet cross-sections in $e^+ e^-$
  annihilation}},  {\em Phys.Lett.} {\bf B269} (1991) 432--438.

\bibitem{Catani:1991pm}
S.~Catani, Y.~L. Dokshitzer, F.~Fiorani, and B.~Webber, {\it {Average number of
  jets in $e^+ e^-$ annihilation}},  {\em Nucl.Phys.} {\bf B377} (1992)
  445--460.

\bibitem{Dissertori:1995qx}
G.~Dissertori and M.~Schmelling, {\it {An Improved theoretical prediction for
  the two jet rate in $e^+ e^-$ annihilation}},  {\em Phys.Lett.} {\bf B361}
  (1995) 167--178.

\bibitem{Banfi:2001bz}
A.~Banfi, G.~Salam, and G.~Zanderighi, {\it {Semi-numerical resummation of
  event shapes}},  {\em JHEP} {\bf 0201} (2002) 018,
  [\href{http://xxx.lanl.gov/abs/hep-ph/0112156}{{\tt hep-ph/0112156}}].

\bibitem{Dasgupta:2003iq}
M.~Dasgupta and G.~P. Salam, {\it {Event shapes in $e^+ e^-$ annihilation and
  deep inelastic scattering}},  {\em J.Phys.G} {\bf G30} (2004) R143,
  [\href{http://xxx.lanl.gov/abs/hep-ph/0312283}{{\tt hep-ph/0312283}}].

\bibitem{GehrmannDeRidder:2007bj}
A.~Gehrmann-De~Ridder, T.~Gehrmann, E.~Glover, and G.~Heinrich, {\it
  {Second-order QCD corrections to the thrust distribution}},  {\em
  Phys.Rev.Lett.} {\bf 99} (2007) 132002,
  [\href{http://xxx.lanl.gov/abs/arXiv:0707.1285}{{\tt arXiv:0707.1285}}].

\bibitem{Dissertori:2007xa}
G.~Dissertori, A.~Gehrmann-De~Ridder, T.~Gehrmann, E.~Glover, G.~Heinrich, {\em
  et~al.}, {\it {First determination of the strong coupling constant using NNLO
  predictions for hadronic event shapes in $e^+ e^-$ annihilations}},  {\em
  JHEP} {\bf 0802} (2008) 040,
  [\href{http://xxx.lanl.gov/abs/arXiv:0712.0327}{{\tt arXiv:0712.0327}}].

\bibitem{Dissertori:2009qa}
G.~Dissertori, A.~Gehrmann-De~Ridder, T.~Gehrmann, E.~Glover, G.~Heinrich, {\em
  et~al.}, {\it {Precise determination of the strong coupling constant at NNLO
  in QCD from the three-jet rate in electron--positron annihilation at LEP}},
  {\em Phys.Rev.Lett.} {\bf 104} (2010) 072002,
  [\href{http://xxx.lanl.gov/abs/arXiv:0910.4283}{{\tt arXiv:0910.4283}}].

\bibitem{Weinzierl:2009ms}
S.~Weinzierl, {\it {Event shapes and jet rates in electron--positron
  annihilation at NNLO}},  {\em JHEP} {\bf 0906} (2009) 041,
  [\href{http://xxx.lanl.gov/abs/arXiv:0904.1077}{{\tt arXiv:0904.1077}}].

\bibitem{Becher:2008cf}
T.~Becher and M.~D. Schwartz, {\it {A Precise determination of $\alpha_s$ from
  LEP thrust data using effective field theory}},  {\em JHEP} {\bf 07} (2008)
  034, [\href{http://xxx.lanl.gov/abs/arXiv:0803.0342}{{\tt arXiv:0803.0342}}].

\bibitem{Chien:2010kc}
Y.-T. Chien and M.~D. Schwartz, {\it {Resummation of heavy jet mass and
  comparison to LEP data}},
  \href{http://xxx.lanl.gov/abs/arXiv:1005.1644}{{\tt arXiv:1005.1644}}.

\bibitem{Sjostrand:2006za}
T.~Sjostrand, S.~Mrenna, and P.~Z. Skands, {\it {PYTHIA 6.4 Physics and
  Manual}},  {\em JHEP} {\bf 0605} (2006) 026,
  [\href{http://xxx.lanl.gov/abs/hep-ph/0603175}{{\tt hep-ph/0603175}}].

\bibitem{Corcella:2000bw}
G.~Corcella, I.~Knowles, G.~Marchesini, S.~Moretti, K.~Odagiri, {\em et~al.},
  {\it {HERWIG 6: An Event generator for hadron emission reactions with
  interfering gluons (including supersymmetric processes)}},  {\em JHEP} {\bf
  0101} (2001) 010, [\href{http://xxx.lanl.gov/abs/hep-ph/0011363}{{\tt
  hep-ph/0011363}}].

\bibitem{Lonnblad:1992tz}
L.~Lonnblad, {\it {ARIADNE version 4: A Program for simulation of QCD cascades
  implementing the color dipole model}},  {\em Comput.Phys.Commun.} {\bf 71}
  (1992) 15--31.

\bibitem{Dokshitzer:1995qm}
Y.~L. Dokshitzer, G.~Marchesini, and B.~Webber, {\it {Dispersive approach to
  power behaved contributions in QCD hard processes}},  {\em Nucl.Phys.} {\bf
  B469} (1996) 93--142, [\href{http://xxx.lanl.gov/abs/hep-ph/9512336}{{\tt
  hep-ph/9512336}}].

\bibitem{Barate:1996fi}
{\bf ALEPH} Collaboration, R.~Barate {\em et~al.}, {\it {Studies of quantum
  chromodynamics with the ALEPH detector}},  {\em Phys.Rept.} {\bf 294} (1998)
  1--165.

\bibitem{Dissertori:2009ik}
G.~Dissertori, A.~Gehrmann-De~Ridder, T.~Gehrmann, E.~Glover, G.~Heinrich, {\em
  et~al.}, {\it {Determination of the strong coupling constant using matched
  NNLO+NLLA predictions for hadronic event shapes in $e^+e^-$ annihilations}},
  {\em JHEP} {\bf 0908} (2009) 036,
  [\href{http://xxx.lanl.gov/abs/arXiv:0906.3436}{{\tt arXiv:0906.3436}}].

\bibitem{Gleisberg:2003xi}
T.~Gleisberg, S.~Hoche, F.~Krauss, A.~Schalicke, S.~Schumann, {\em et~al.},
  {\it {SHERPA 1.$\alpha$: A Proof of concept version}},  {\em JHEP} {\bf 0402}
  (2004) 056, [\href{http://xxx.lanl.gov/abs/hep-ph/0311263}{{\tt
  hep-ph/0311263}}].

\bibitem{Gleisberg:2008ta}
T.~Gleisberg, S.~Hoche, F.~Krauss, M.~Schonherr, S.~Schumann, {\em et~al.},
  {\it {Event generation with SHERPA 1.1}},  {\em JHEP} {\bf 0902} (2009) 007,
  [\href{http://xxx.lanl.gov/abs/arXiv:0811.4622}{{\tt arXiv:0811.4622}}].

\bibitem{Catani:2001cc}
S.~Catani, F.~Krauss, R.~Kuhn, and B.~Webber, {\it {QCD matrix elements +
  parton showers}},  {\em JHEP} {\bf 0111} (2001) 063,
  [\href{http://xxx.lanl.gov/abs/hep-ph/0109231}{{\tt hep-ph/0109231}}].

\bibitem{Ellis:2007br}
R.~Ellis, W.~Giele, and Z.~Kunszt, {\it {A Numerical Unitarity Formalism for
  Evaluating One-Loop Amplitudes}},  {\em JHEP} {\bf 0803} (2008) 003,
  [\href{http://xxx.lanl.gov/abs/arXiv:0708.2398}{{\tt arXiv:0708.2398}}].

\bibitem{Giele:2008ve}
W.~T. Giele, Z.~Kunszt, and K.~Melnikov, {\it {Full one-loop amplitudes from
  tree amplitudes}},  {\em JHEP} {\bf 0804} (2008) 049,
  [\href{http://xxx.lanl.gov/abs/arXiv:0801.2237}{{\tt arXiv:0801.2237}}].

\bibitem{Ellis:2008qc}
R.~Ellis, W.~Giele, Z.~Kunszt, K.~Melnikov, and G.~Zanderighi, {\it {One-loop
  amplitudes for W+3 jet production in hadron collisions}},  {\em JHEP} {\bf
  0901} (2009) 012, [\href{http://xxx.lanl.gov/abs/arXiv:0810.2762}{{\tt
  arXiv:0810.2762}}].

\bibitem{Frederix:2009yq}
R.~Frederix, S.~Frixione, F.~Maltoni, and T.~Stelzer, {\it {Automation of
  next-to-leading order computations in QCD: The FKS subtraction}},  {\em JHEP}
  {\bf 0910} (2009) 003, [\href{http://xxx.lanl.gov/abs/arXiv:0908.4272}{{\tt
  arXiv:0908.4272}}].

\bibitem{Bern:1994fz}
Z.~Bern, L.~J. Dixon, and D.~A. Kosower, {\it {One loop corrections to two
  quark three gluon amplitudes}},  {\em Nucl.Phys.} {\bf B437} (1995) 259--304,
  [\href{http://xxx.lanl.gov/abs/hep-ph/9409393}{{\tt hep-ph/9409393}}].

\bibitem{DelDuca:1999rs}
V.~Del~Duca, L.~J. Dixon, and F.~Maltoni, {\it {New color decompositions for
  gauge amplitudes at tree and loop level}},  {\em Nucl.Phys.} {\bf B571}
  (2000) 51--70, [\href{http://xxx.lanl.gov/abs/hep-ph/9910563}{{\tt
  hep-ph/9910563}}].

\bibitem{Kniehl:1989bb}
B.~A. Kniehl and J.~H. Kuhn, {\it {QCD Corrections to the Axial Part of the Z
  Decay Rate}},  {\em Phys.Lett.} {\bf B224} (1989) 229.

\bibitem{Hagiwara:1990dx}
K.~Hagiwara, T.~Kuruma, and Y.~Yamada, {\it {Three jet distributions from the
  one-loop Zgg vertex at $e^+ e^-$ colliders}},  {\em Nucl.Phys.} {\bf B358}
  (1991) 80--96.

\bibitem{Berger:2008sz}
C.~F. Berger {\em et~al.}, {\it {One-Loop Multi-Parton Amplitudes with a Vector
  Boson for the LHC}},  \href{http://xxx.lanl.gov/abs/0808.0941}{{\tt
  0808.0941}}.

\bibitem{Berger:2009ep}
C.~F. Berger {\em et~al.}, {\it {Next-to-Leading Order QCD Predictions for
  W+3-Jet Distributions at Hadron Colliders}},  {\em Phys. Rev.} {\bf D80}
  (2009) 074036, [\href{http://xxx.lanl.gov/abs/0907.1984}{{\tt 0907.1984}}].

\bibitem{Berger:2010vm}
C.~F. Berger {\em et~al.}, {\it {Next-to-Leading Order QCD Predictions for
  $Z,\gamma^*+3$~jet Distributions at the Tevatron}},
  \href{http://xxx.lanl.gov/abs/arXiv:1004.1659}{{\tt arXiv:1004.1659}}.

\bibitem{Frixione:1995ms}
S.~Frixione, Z.~Kunszt, and A.~Signer, {\it {Three jet cross-sections to
  next-to-leading order}},  {\em Nucl.Phys.} {\bf B467} (1996) 399--442,
  [\href{http://xxx.lanl.gov/abs/hep-ph/9512328}{{\tt hep-ph/9512328}}].

\bibitem{Alwall:2007st}
J.~Alwall, P.~Demin, S.~de~Visscher, R.~Frederix, M.~Herquet, {\em et~al.},
  {\it {MadGraph/MadEvent v4: The New Web Generation}},  {\em JHEP} {\bf 0709}
  (2007) 028, [\href{http://xxx.lanl.gov/abs/arXiv:0706.2334}{{\tt
  arXiv:0706.2334}}].

\bibitem{Binoth:2010xt}
T.~Binoth, F.~Boudjema, G.~Dissertori, A.~Lazopoulos, A.~Denner, {\em et~al.},
  {\it {A Proposal for a standard interface between Monte Carlo tools and
  one-loop programs}},  {\em Comput.Phys.Commun.} {\bf 181} (2010) 1612--1622,
  [\href{http://xxx.lanl.gov/abs/arXiv:1001.1307}{{\tt arXiv:1001.1307}}].

\bibitem{Maltoni:2002qb}
F.~Maltoni and T.~Stelzer, {\it {MadEvent: Automatic event generation with
  MadGraph}},  {\em JHEP} {\bf 0302} (2003) 027,
  [\href{http://xxx.lanl.gov/abs/hep-ph/0208156}{{\tt hep-ph/0208156}}].

\bibitem{Andersson:1983ia}
B.~Andersson, G.~Gustafson, G.~Ingelman, and T.~Sjostrand, {\it {Parton
  Fragmentation and String Dynamics}},  {\em Phys.Rept.} {\bf 97} (1983)
  31--145.

\bibitem{Winter:2003tt}
J.-C. Winter, F.~Krauss, and G.~Soff, {\it {A Modified cluster hadronization
  model}},  {\em Eur.Phys.J.} {\bf C36} (2004) 381--395,
  [\href{http://xxx.lanl.gov/abs/hep-ph/0311085}{{\tt hep-ph/0311085}}].

\bibitem{Martin:2009iq}
A.~Martin, W.~Stirling, R.~Thorne, and G.~Watt, {\it {Parton distributions for
  the LHC}},  {\em Eur.Phys.J.} {\bf C63} (2009) 189--285,
  [\href{http://xxx.lanl.gov/abs/arXiv:0901.0002}{{\tt arXiv:0901.0002}}].

\bibitem{pdg}
{\bf Particle Data Group} Collaboration, C.~Amsler {\em et~al.}, {\it {Review
  of Particle Physics}},  {\em Phys.Lett.} {\bf B667} (2008) 1. See most recent
  and 2009 partial update for the 2010 edition.

\bibitem{Bethke:2009jm}
S.~Bethke, {\it {The 2009 World Average of $\alpha_s$}},  {\em Eur.Phys.J.}
  {\bf C64} (2009) 689--703,
  [\href{http://xxx.lanl.gov/abs/arXiv:0908.1135}{{\tt arXiv:0908.1135}}].

\bibitem{Jones:2006nq}
R.~Jones, {\it {Final $\alpha_s$ combinations from the LEP QCD Working Group}},
   {\em Nucl.Phys.Proc.Suppl.} {\bf 152} (2006) 15--22.

\bibitem{Jones:2003yv}
R.~W.~L. Jones, M.~Ford, G.~P. Salam, H.~Stenzel, and D.~Wicke, {\it
  {Theoretical uncertainties on $\alpha_s$ from event-shape variables in $e^+
  e^-$ annihilations}},  {\em JHEP} {\bf 12} (2003) 007,
  [\href{http://xxx.lanl.gov/abs/hep-ph/0312016}{{\tt hep-ph/0312016}}].

\bibitem{Brambilla:2007cz}
N.~Brambilla, X.~Garcia~i Tormo, J.~Soto, and A.~Vairo, {\it {Extraction of
  $\alpha_s$ from radiative $\Upsilon(1S)$ decays}},  {\em Phys. Rev.} {\bf
  D75} (2007) 074014, [\href{http://xxx.lanl.gov/abs/hep-ph/0702079}{{\tt
  hep-ph/0702079}}].

\bibitem{Glasman:2007sm}
{\bf H1} Collaboration, C.~Glasman, {\it {Precision measurements of $\alpha_s$
  at HERA}},  {\em J. Phys. Conf. Ser.} {\bf 110} (2008) 022013,
  [\href{http://xxx.lanl.gov/abs/arXiv:0709.4426}{{\tt arXiv:0709.4426}}].

\bibitem{Blumlein:2007dk}
J.~Blumlein, {\it {$\Lambda_{\rm QCD}$ and $\alpha_s(M_Z^2)$ from DIS Structure
  Functions}},  \href{http://xxx.lanl.gov/abs/arxiv:0706.2430}{{\tt
  arxiv:0706.2430}}.

\bibitem{Abbate:2010xh}
R.~Abbate, M.~Fickinger, A.~H. Hoang, V.~Mateu, and I.~W. Stewart, {\it {Thrust
  at N${}^3$LL with Power Corrections and a Precision Global Fit for
  $\alpha_s(m_Z)$}},  \href{http://xxx.lanl.gov/abs/arXiv:1006.3080}{{\tt
  arXiv:1006.3080}}.

\bibitem{Davies:2008sw}
{\bf HPQCD} Collaboration, C.~T.~H. Davies {\em et~al.}, {\it {Update: Accurate
  Determinations of $\alpha_s$ from Realistic Lattice QCD}},  {\em Phys. Rev.}
  {\bf D78} (2008) 114507, [\href{http://xxx.lanl.gov/abs/arXiv:0807.1687}{{\tt
  arXiv:0807.1687}}].

\bibitem{Flacher:2008zq}
H.~Flacher {\em et~al.}, {\it {Gfitter - Revisiting the Global Electroweak Fit
  of the Standard Model and Beyond}},  {\em Eur. Phys. J.} {\bf C60} (2009)
  543--583, [\href{http://xxx.lanl.gov/abs/arXiv:0811.0009}{{\tt
  arXiv:0811.0009}}].

\end{thebibliography}\endgroup
\bibliographystyle{JHEP}

\end{document}